\documentclass[usenatbib]{mn2e}
\usepackage{psfig}

\def\gs{\mathrel{\raise0.35ex\hbox{$\scriptstyle >$}\kern-0.6em
\lower0.40ex\hbox{{$\scriptstyle \sim$}}}}
\def\ls{\mathrel{\raise0.35ex\hbox{$\scriptstyle <$}\kern-0.6em
\lower0.40ex\hbox{{$\scriptstyle \sim$}}}}

\def\b59{MIPS\,J142824.0+352619} %%{B{\" o}{\" o}tes.59}
\def\kms{\,\hbox{km}\,\hbox{s}^{-1}}
\def\Msol{\mathrel{\rm M_{\odot}}}
\def\Lsol{\mathrel{\rm L_{\odot}}}
\def\Msolyr{\mathrel{\rm M_{\odot}yr^{-1}}}
\def\Wm2{\,\hbox{W}\,\hbox{m}^{-2}}
\def\gsim{\mathrel{\raise0.35ex\hbox{$\scriptstyle >$}\kern-0.6em
\lower0.40ex\hbox{{$\scriptstyle \sim$}}}}
\def\lsim{\mathrel{\raise0.35ex\hbox{$\scriptstyle <$}\kern-0.6em
\lower0.40ex\hbox{{$\scriptstyle \sim$}}}}

\begin{document}

\title[Near-Infrared IFU Spectroscopy of SCUBA Galaxies]{The Link
  between Submillimetre Galaxies and Luminous Ellipticals: Near-Infrared IFU
  Spectroscopy of Submillimetre Galaxies}

\author[Swinbank et al.]
{ \parbox[h]{\textwidth}{ 
A.\,M.\ Swinbank,$^{\! 1\, *}$
S.\,C.\ Chapman,$^{\! 2}$ 
Ian Smail,$^{\!1}$
C.\, Lindner,$^{\! 2}$ 
C.\, Borys,$^{\! 2,3}$
A.\,W.\ Blain,$^{\! 2}$
R.\,J.\ Ivison,$^{4,5}$
\& G.\,F.\ Lewis$^{6}$}
\vspace*{6pt} \\
$^1$Institute for Computational Cosmology, Department of
Physics, Durham University, South Road, Durham, DH1 3LE, UK \\
$^2$Astronomy Department, California Institute of
Technology, 105-24, Pasadena, CA 91125, USA \\
$^3$University of Toronto,  Toronto, Ontario, Canada M5S 3H8, Canada \\
$^4$Astronomy Technology Center, Royal Observatory,
Blackford Hill, Edinburgh, EH19 3HJ, UK \\
$^5$Institute for Astronomy, University of Edinburgh,
Edinburgh, EH19 3HJ, UK \\
$^6$School of Physics, University of Sydney, NSW 2006, Australia \\
$^*$Email: a.m.swinbank@durham.ac.uk \\
}
\maketitle
\begin{abstract}
  We present two-dimensional spectroscopy covering the rest-frame
  wavelengths of strong optical emission lines in six luminous
  submillimetre galaxies at $z=1.3$--2.5.  Using this near-infrared
  integral field spectroscopy together with {\it Hubble Space
    Telescope} ACS and NICMOS imaging we map the dynamics and
  morphologies of these systems on scales from 4--11\,kpc.  Four of the
  systems show multiple components in their spatially-resolved spectra
  with average velocity offsets of $\sim$\,180$\kms$ across 8\,kpc in
  projection.  From the ensemble properties of eight galaxies, from our
  survey and the literature, we estimate the typical dynamical masses
  of bright submillimetre galaxies as 5$\pm$3$\times$10$^{11}\Msol$.
  This is similar to recent estimates of their stellar masses --
  suggesting that the dynamics of the central regions of these galaxies
  are baryon dominated, with a substantial fraction of those baryons in
  stars by the epoch of observation.  Combining our dynamical mass
  estimates with stellar luminosities for this population we
  investigate whether submillimetre galaxies can evolve onto the
  Faber-Jackson relation for local ellipticals.  Adopting a typical
  lifetime of $\tau_{\rm burst}\sim 300$\,Myr for the
  submillimetre-luminous phase -- using the latest estimates of gas
  masses, star formation rates and AGN contribution to the bolometric
  luminosities -- we find that the stellar populations of submillimetre
  galaxies should fade to place them on the Faber-Jackson relation, at
  M$_K\sim -25.1$.  Furthermore, using the same starburst lifetime we
  correct the observed space density of submillimetre galaxies for the
  duty cycle to derive a volume density of the progenitors of
  $\sim1\times10^{-4}$\,Mpc$^{-3}$.  This is consistent with the space
  density of local luminous early-type galaxies with M$_K\sim -25.1$,
  indicating that submillimetre galaxies can evolve onto the scaling
  relations observed for local early-type galaxies, and the observed
  population at $z\sim2$ is then sufficient to account for the
  formation of the whole population of $\gsim3 L^*_K$ ellipticals seen at $z\sim 0$.
\end{abstract}


\section{Introduction}

Deep optical and near-infrared imaging with the superlative resolution
of the NICMOS and ACS cameras on-board {\it Hubble Space Telescope}
({\it HST}) has made it possible to study the morphologies and colours
of high-redshift far-infrared luminous galaxies identified in the
sub-millimetre (submm) waveband in unprecedented levels of detail
\citep{Smail98,Smail02,Chapman03b,Smail04,Pope05,Almaini05}.  By
combining this data with spectroscopically confirmed redshifts
\citep{Chapman03a,Chapman05a}, we are beginning to understand the
processes which triggers the immense bolometric luminosity output in
these galaxies, allowing us to address issues such as their true
contribution to the star-formation rate history of the Universe
\citep[e.g.,][]{Chapman05a}.  Submm galaxies (SMGs) are amongst the
most bolometrically luminous galaxies in the Universe (with $L_{\rm
  bol}\sim$10$^{13}\Lsol$ and a median redshift of $<z>\sim$2.2).
However, in contrast to quasars at the same early epochs, their
bolometric output appears to be dominated by star-formation rather than
AGN activity \citep{Swinbank04,Alexander05a}.  It is clearly important
to establish firmly the masses for these galaxies in order to
understanding the rapid evolution of the submm population.  Such
diagnostics will allow us to determine how they relate to present-day
galaxies and in particular to test whether they represent the formation
phase of the most massive elliptical galaxies at the present-day, as
many suspect \citep{Lilly99}.

To probe the dynamical structures of these frequently complex systems
requires a reliable separation of the spatial and spectral information.
One particularly powerful approach to achieve this is using millimetric
CO emission line maps to trace the distribution and kinematics of dense
gas within the galaxies \citep{Frayer99,Frayer03,Neri03,Greve05}
especially at high resolution \citep{Genzel03,Downes03,Tacconi06}.
Unfortunately, prior to the commissioning of the Atacama Large
Millimetre Array, this approach remains observationally very demanding,
typically requiring 30 hours per source with the IRAM interferometer
and can only be applied to the most massive and gas-rich SMGs.  An
alternative tool which is less demanding in terms of telescope time and
can provide spatially-resolved dynamics of SMGs with sufficient spatial
and velocity resolution to study their internal structure is
near-infrared integral field spectroscopy targeting the rest-frame
optical emission lines such as [O{\sc iii}]$\lambda$5007 and
H$\alpha\lambda$6563 \citep[e.g.\,][]{Tecza04,Swinbank05b}.  The
spectroscopic maps so produced allow us trace the dynamical and
structural properties of SMGs on scales of a few kpc, pin-point the
sites of active star formation and identify non-thermal emission from
active galactic nuclei (AGN) in components within these systems.
However, they are also subject to potential biases due to extreme
extinction within some regions of the SMGs and from outflows in the
emission-line gas which is being observed \citep{Bower04,Wilman05}.
While the star formation rates estimated from the H$\alpha$ are
typically a factor of 10--100\,$\times$ less than that predicted from
the far-infrared emission (suggesting to 1--4 magnitudes of dust
absorption; q.v. \citealt{Smail04}), it is clear that the H$\alpha$
emission provides a more secure probe of the star-formation rates and
dynamics than either the Ly$\alpha$ emission or rest-frame UV continuum
and can thus provide a reliable and observationally efficient probe of
the activity and dynamics of this important population.

Using the UIST near-infrared Integral Field Unit (IFU) at UKIRT and the
GNIRS IFU at Gemini-South we have studied the rest-frame optical
emission line properties of six submillimetre galaxies at $z=1.3$--2.5.
These observations, together with high-resolution imaging from {\it
  HST} allow us to probe the physical cause of the rapid evolution of
submillimetre galaxies, enabling us to test whether these far-infrared
luminous galaxies comprise merging systems which are likely to be the
progenitors of local massive ellipticals, or whether instead they are
simply high-luminosity episodes in the history of more mundane
galaxies. Our paper is laid out as follows: in \S\ref{sec:obs_red} we
present the observations and their reduction.  In \S\ref{sec:analysis}
we discuss the analysis of these data, while in \S\ref{sec:discussion}
and \S\ref{sec:conc} we present our discussion and conclusions
respectively.  We use a cosmology with $H_{0}=70\kms$, $\Omega_{M}=0.3$
and $\Omega_{\Lambda}=0.7$ in which 1$''$ corresponds to 8.2\,kpc at
$z=2.5$ and 8.5\,kpc at $z=1.5$.

%%%%%%%%%%%%%%%%%%%%%%%%%%%%%%%%%%%%%%%%%%%%%%%%%%%%%%%%%%%%%%%%%%%%%%%%%%%%%

\begin{figure*}

\begin{minipage}{7.5in}
\centerline{\psfig{figure=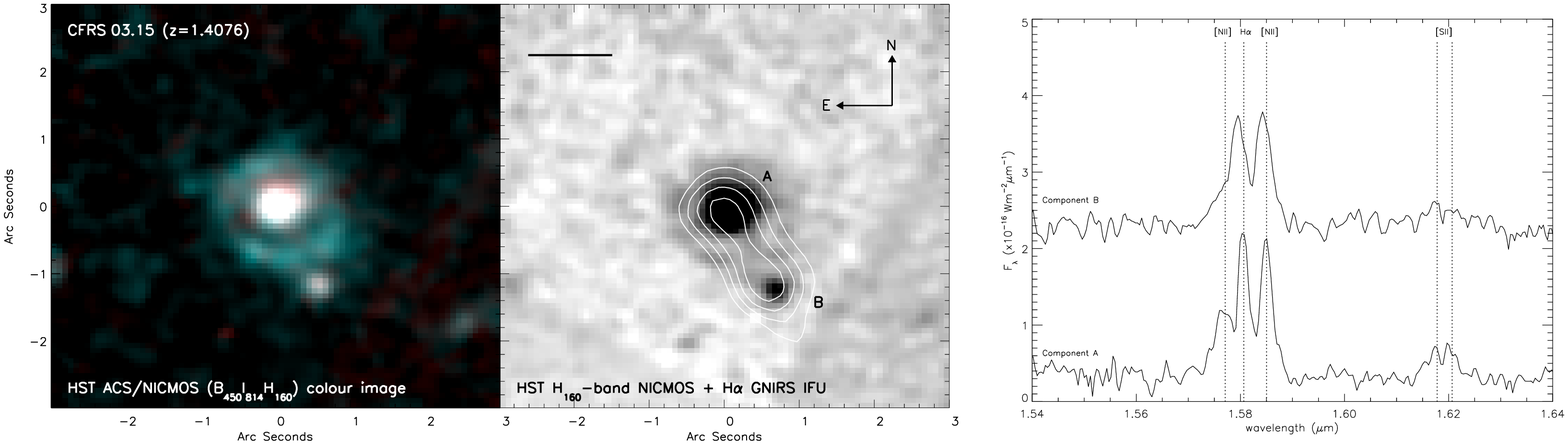,width=6.5in,angle=0}}
\end{minipage}
\begin{minipage}{7.5in}
{\addtolength{\baselineskip}{-3pt}
  {\scriptsize
 \noindent{\bf Figure~1.} {\it Left:} True colour {\it HST} ACS/NICMOS
 $B_{450}I_{814}H_{160}$-band image of CFRS\,03.15 at $z$=1.4076
 showing the complex morphology (the colour image is displayed in log
 scale to emphasise the morphology).  {\it Middle:} {\it HST} NICMOS
 $H_{160}$-band image of CFRS\,03.15 with the H$\alpha$ emission line
 map from the GNIRS IFU observations overlaid as contours, showing that
 the bright knot located $\sim$1.3$''$ ($\sim$11\,kpc) to the
 South-West is associated with the system.  The solid bar represents
 the FWHM of the seeing disk for the IFU observations.  {\it Right:}
 Spectra from the two components seen in the {\it HST} imaging of
 CFRS\,03.15 from the Keck/NIRSPEC observation.  The top spectrum is
 offset in flux for clarity.  From our spectroscopy, the compact
 component to the South-West is blue-shifted by 90$\pm$20$\kms$ in
 projection from the nucleus, suggesting they are two components whose
 interaction has promoted the strong starburst and disturbed
 morphology.  It is interesting to note that both spectra have strong
 [N{\sc iii}]/H$\alpha$ emission line flux ratios which may indicate
 that both components host active AGN.

}}
\end{minipage}

\begin{minipage}{7.5in}
\centerline{
\psfig{figure=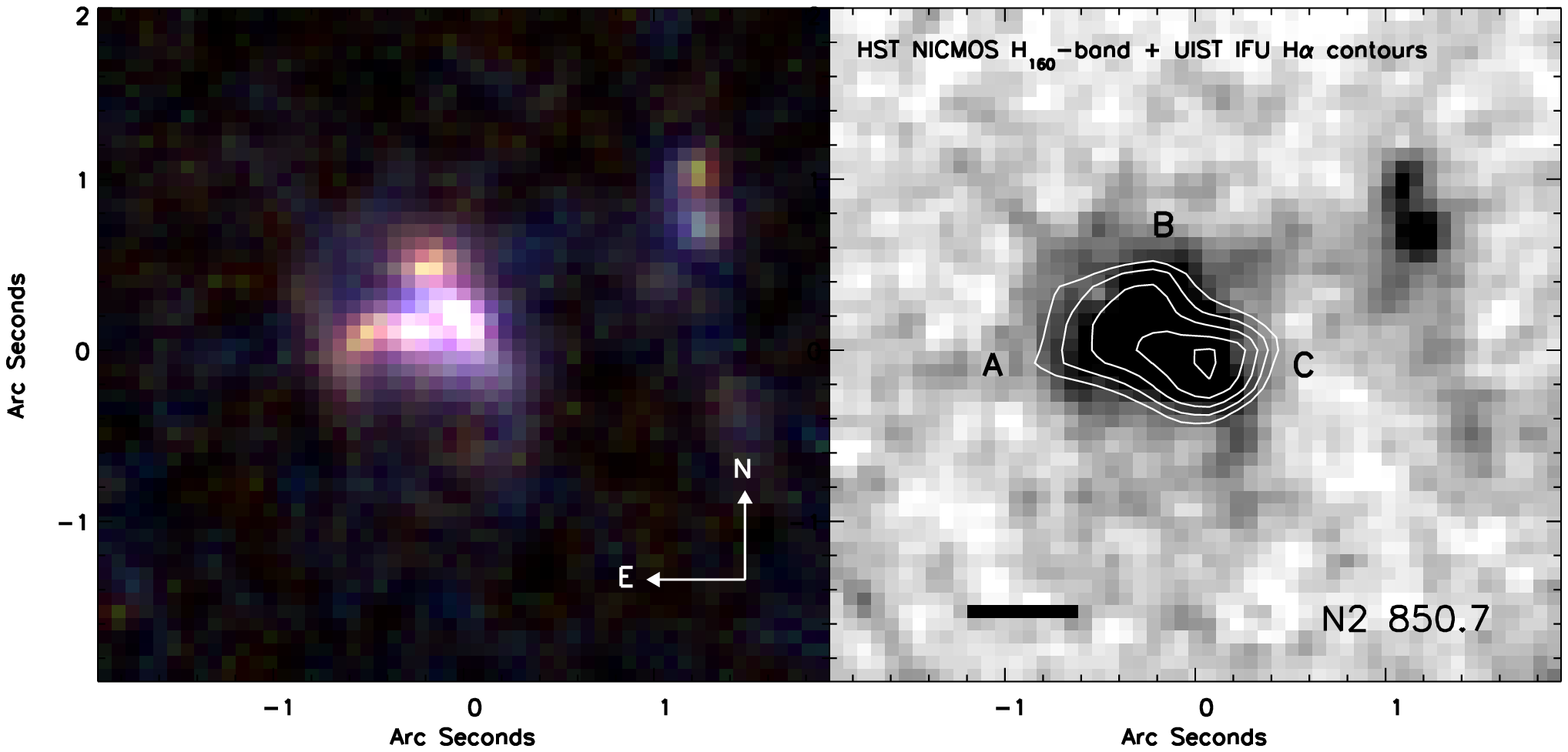,width=4.0in,angle=90}
\psfig{figure=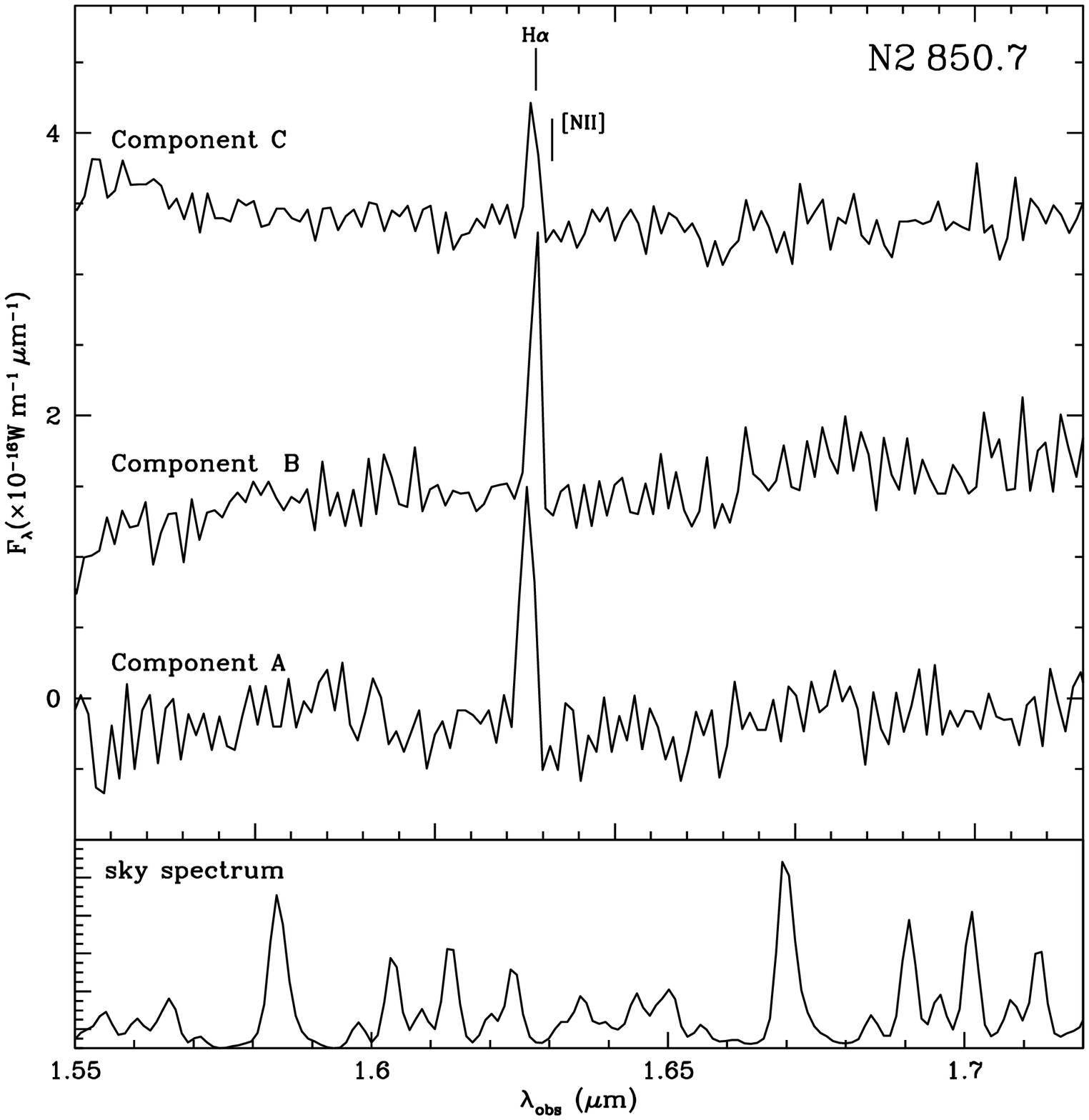,angle=0,width=2.3in}}
\end{minipage}
\begin{minipage}{7.5in}
{\addtolength{\baselineskip}{-3pt}
{\scriptsize
  \noindent{\bf Figure~2.}  {\it Left:} A true colour image of N2\,850.7
  comprising {\it HST} ACS $I_{814}$-band and NICMOS $H_{160}$-band
  images illustrating the complex morphology of this system.  {\it
    Middle:} NICMOS $H_{160}$-band image with H$\alpha$ emission line
  intensity overlaid as contours.  Each panel is 4\,arcsec square
  (34\,kpc at $z=1.49$) and has East left and North top.  The bar shown
  on the NICMOS image indicates the 0.6$''$ resolution of the IFU
  observations.  The system comprises at least three components, marked
  as A, B and C (and there is also tentative evidence for weak
  H$\alpha$ emission from the source $\sim1''$ to the West, suggesting
  it may be associated with the system).  The continuum colours of C
  are bluer than the other components.  The intensity of the H$\alpha$
  emission traces the $H_{160}$-band morphology and it is interesting
  to note that the star formation appears distributed across the SMG.
  {\it Right:} Spectra covering the redshifted H$\alpha$ emission in
  the three components of N2\,850.7 from the UIST IFU observations.
  The spectra have been offset in flux scale for clarity.  The lower
  panel shows a sky spectrum.  By centroiding the emission lines, we
  find velocity offsets between components A \& B and A \& C of
  $215\pm80\kms$ and $135\pm90\kms$ in projection respectively.

}}
\end{minipage}
\vspace{0.5cm}

\begin{minipage}{7.5in}
\centerline{
\psfig{figure=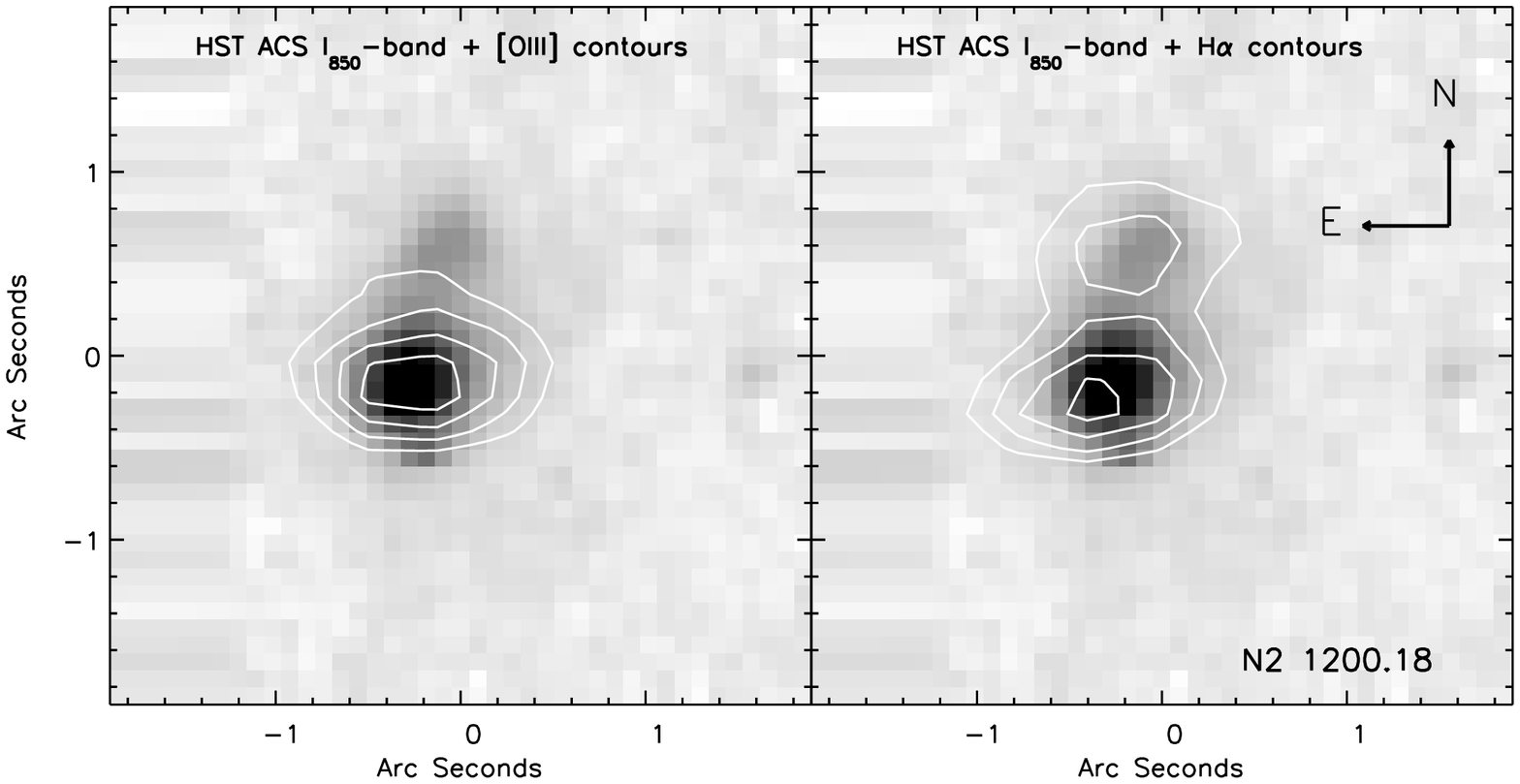,width=3.5in,angle=90}
\psfig{figure=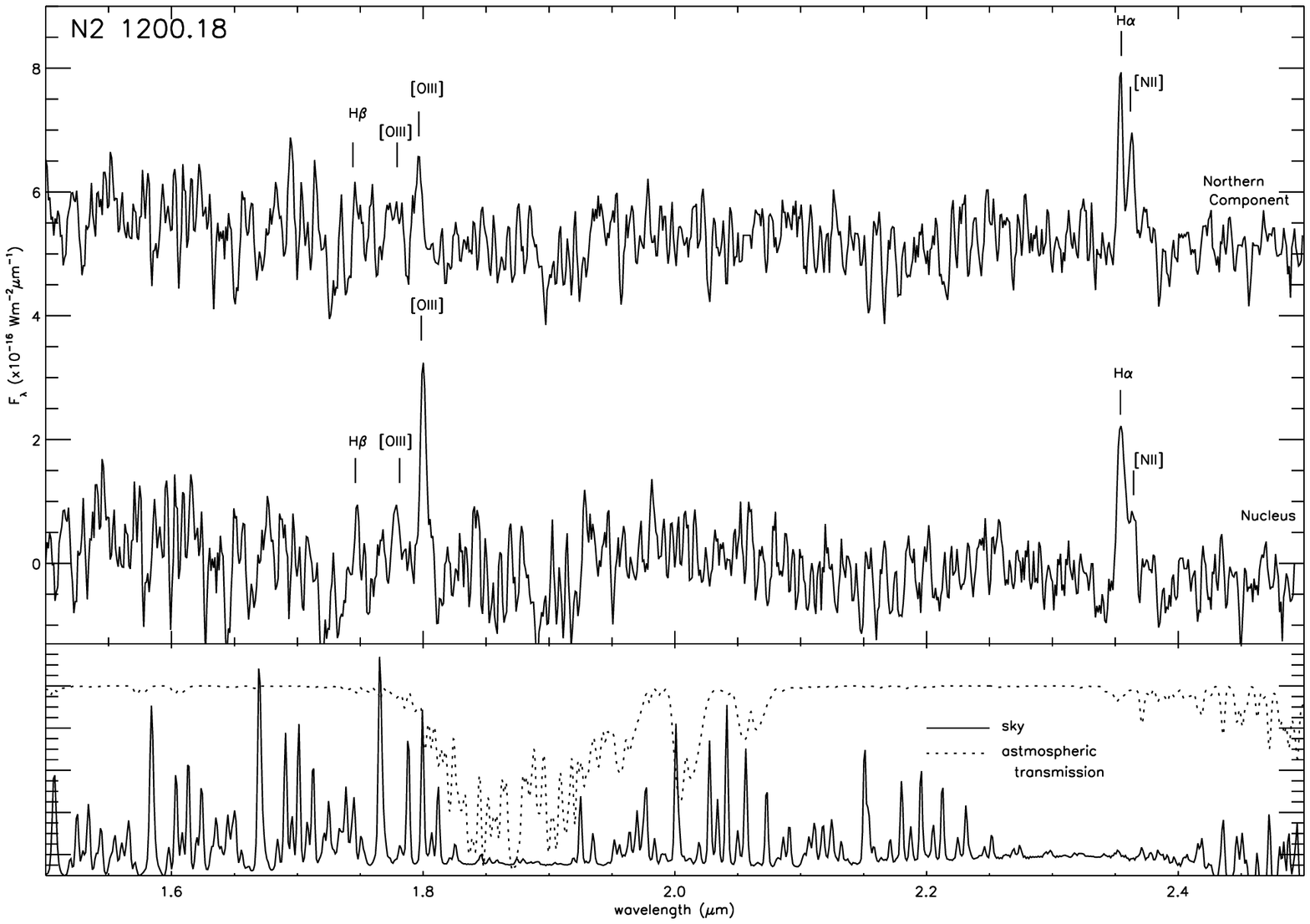,angle=90,width=3.9in}}
\end{minipage}
\begin{minipage}{7.5in}
{\addtolength{\baselineskip}{-3pt}
{\scriptsize
  \noindent{\bf Figure~3.}  {\it Left:} {\it HST} $I_{814}$-band image
  of N2\,1200.18 with the [O{\sc iii}]$\lambda$5007 and H$\alpha$
  emission line intensity maps from the UIST IFU observations overlaid
  as contours. Both panels are displayed on a $\log$ scale to emphasis
  the morphology.  The {\it HST} image shows a a central bright,
  compact component and a lower-surface brightness extension to the
  North.  The [O{\sc iii}]$\lambda$5007 and H$\alpha$ emission also
  have different morphologies, with the strong, compact [O{\sc
    iii}]$\lambda$5007 probably arising from (unresolved) AGN activity
  in the bright central component, whilst the H$\alpha$ morphology
  appears resolved and extends across the whole system.  {\it Right:}
  Spectra from the central component and northern extension of
  N2\,1200.18 around the [O{\sc iii}]$\lambda$5007 and H$\alpha$
  emission lines from the UIST IFU observations.  The spectrum of the
  central source has an [O{\sc iii}]/H$\alpha$ emission line flux ratio
  of 2.0$\pm$0.5 which, combined with the unresolved [O{\sc iii}]
  emission suggests AGN activity.  The northern component has a
  velocity offset of 250$\pm$80$\kms$ from the nucleus.  This northern
  component has star-burst (rather than AGN) characteristics, with a
  lower [O{\sc iii}]/H$\alpha$ and [N{\sc ii}]/H$\alpha$ emission line
  flux ratios.  {\it Bottom:} The OH airglow emission from the sky
  (continuous line), as well as the Mauna Kea atmospheric transmission
  (dotted line) are also shown in the lower panel.

}}
\end{minipage}
%\vspace{0.3cm}

\end{figure*}

\begin{figure*}
\begin{minipage}{7.5in}
  \centerline{\psfig{figure=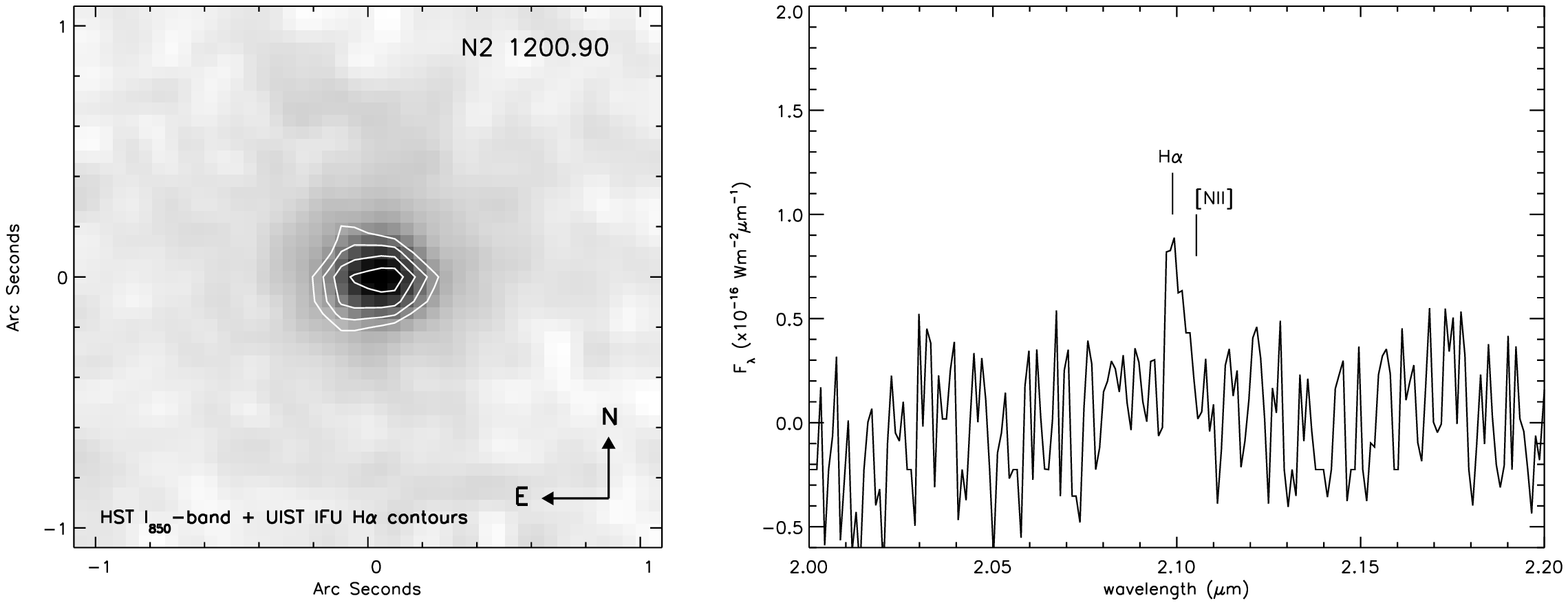,width=6in,angle=90}}
\end{minipage}
\begin{minipage}{7.5in}
{\addtolength{\baselineskip}{-3pt}
{\scriptsize
  \noindent{\bf Figure~4.}  {\it Left:} {\it HST} I$_{814}$-band image
  of N2\,1200.90.  The galaxy has a compact morphology ($\ls$0.2$''$
  (1.6\,kpc) FWHM) in this continuum imaging.  {\it Right:} Spectrum of
  N2\,1200.90 from the UIST IFU observations showing H$\alpha$ at a
  redshift of $z=2.198$. The rest-frame UV spectrum from Chapman et al.\
  (2006) shows clear signs of AGN activity, yet the apparently low
  [N{\sc ii}]/H$\alpha$ emission line ratio from our near-infrared
  spectroscopy is more suggestive of star-formation.  However, we note
  that the H$\alpha$ emission line width of 500$\pm$100$\kms$ may indicate
  (non-thermal) AGN activity.
  
}}
\end{minipage}
\vspace{0.6cm}

\begin{minipage}{7.5in}
\centerline{
\psfig{figure=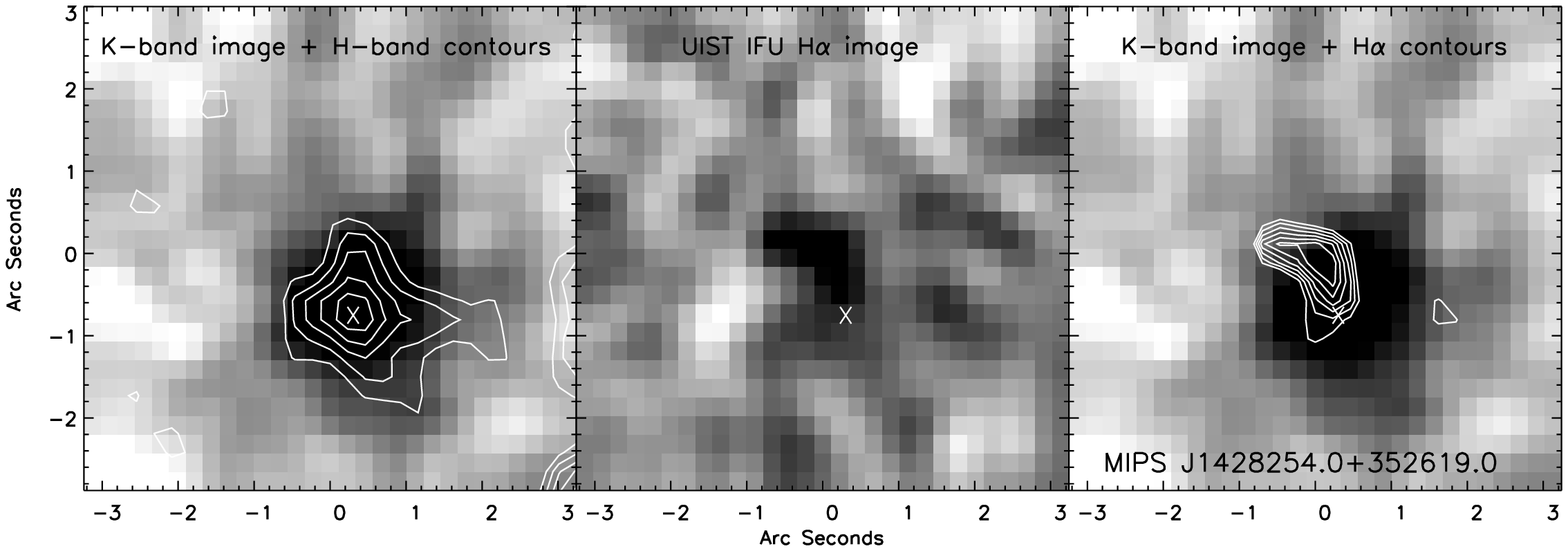,width=5.2in,angle=90}
\psfig{figure=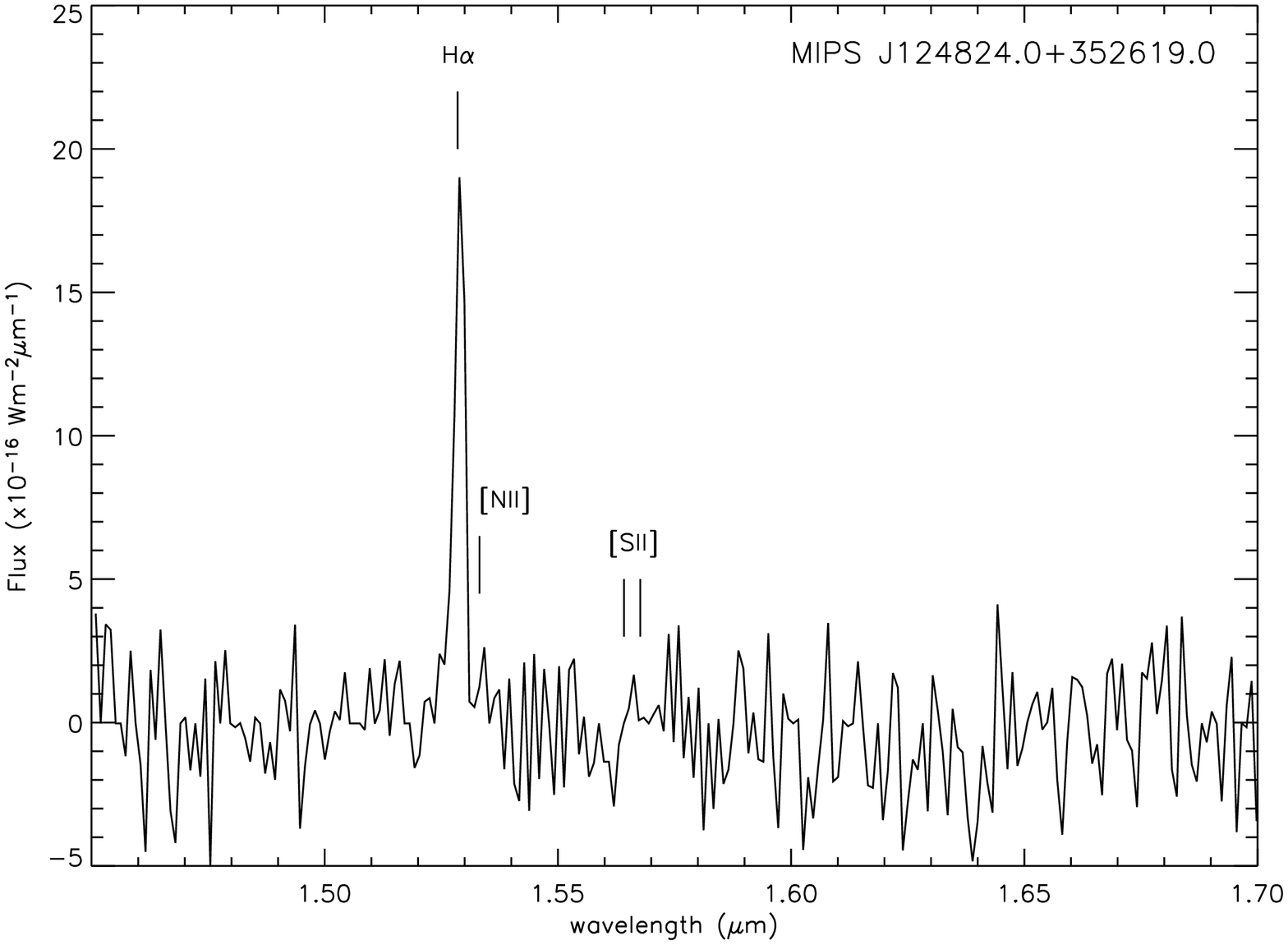,width=2.4in,angle=90}}
\end{minipage}
\begin{minipage}{7.5in}
{\addtolength{\baselineskip}{-3pt}
{\scriptsize
  \noindent{\bf Figure~5.}  Images of \b59 generated from the UIST IFU
  observations.  {\it Left:} $K$-band continuum image (collapsed from
  the datacube between 2.0 and 2.4$\mu$m) with contours overlaid from
  the $H$-band (collapsed between 1.5 to 1.8$\mu$m). {\it Middle:}
  Continuum subtracted H$\alpha$ image. {\it Right:} $K$-band image
  with contours from the continuum subtracted, H$\alpha$ narrow-band
  image overlaid.  The ``X'' in each panel marks the position of the
  center of the $H$-band light distribution.  In the final panel we
  show the collapsed, one-dimensional spectrum around the redshifted
  H$\alpha$ emission from \b59.  The H$\alpha$ redshift of
  $z$=1.328$\pm$0.003 is in excellent agreement with the CO(5-4)
  emission, and the H$\alpha$ line width and emission line flux are
  also comparable to those given in Borys et al.\ (2006).

}}
\end{minipage}
\vspace{0.6cm}

\begin{minipage}{7.5in}
\centerline{
\psfig{figure=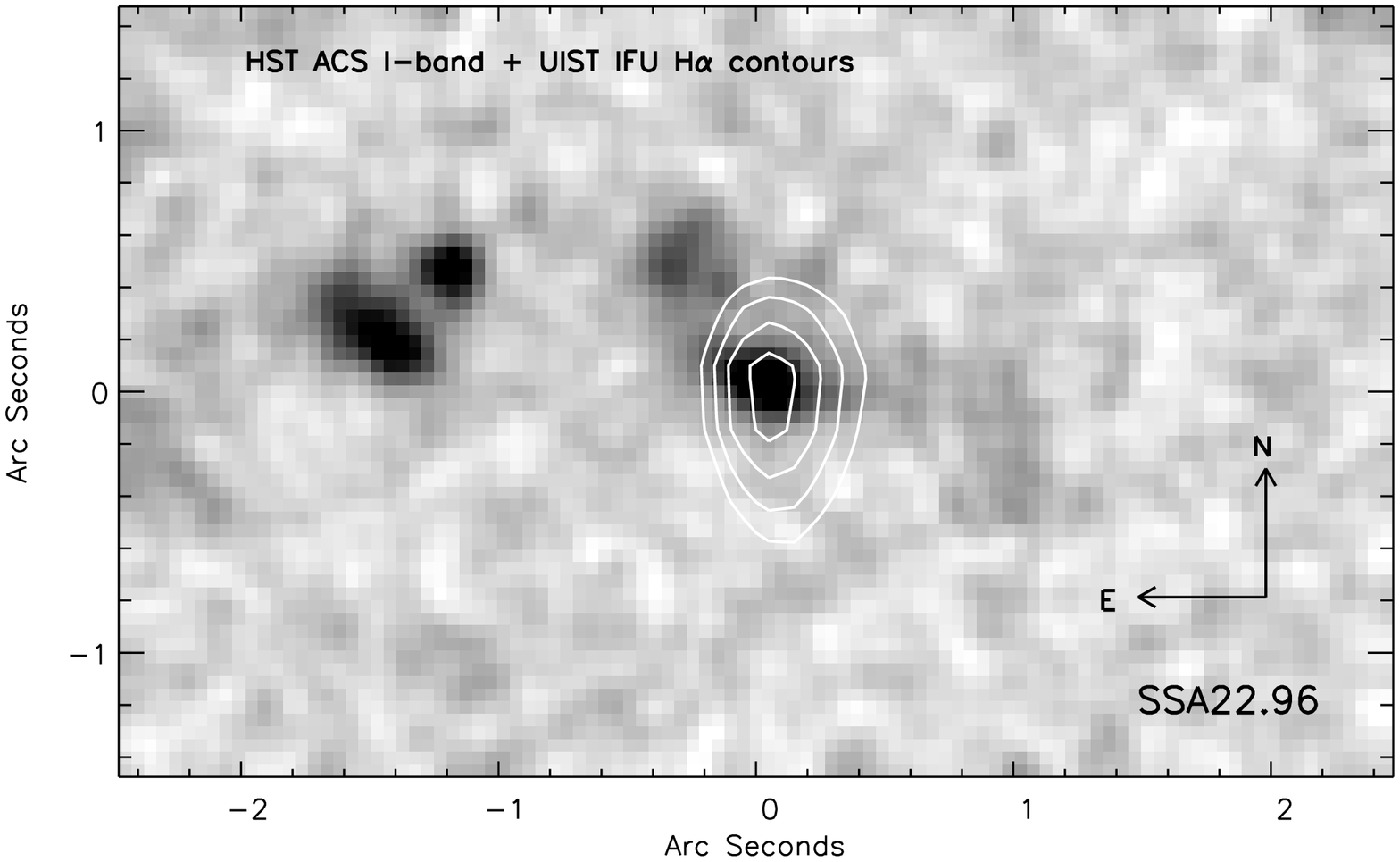,width=3.0in,angle=90}
\psfig{figure=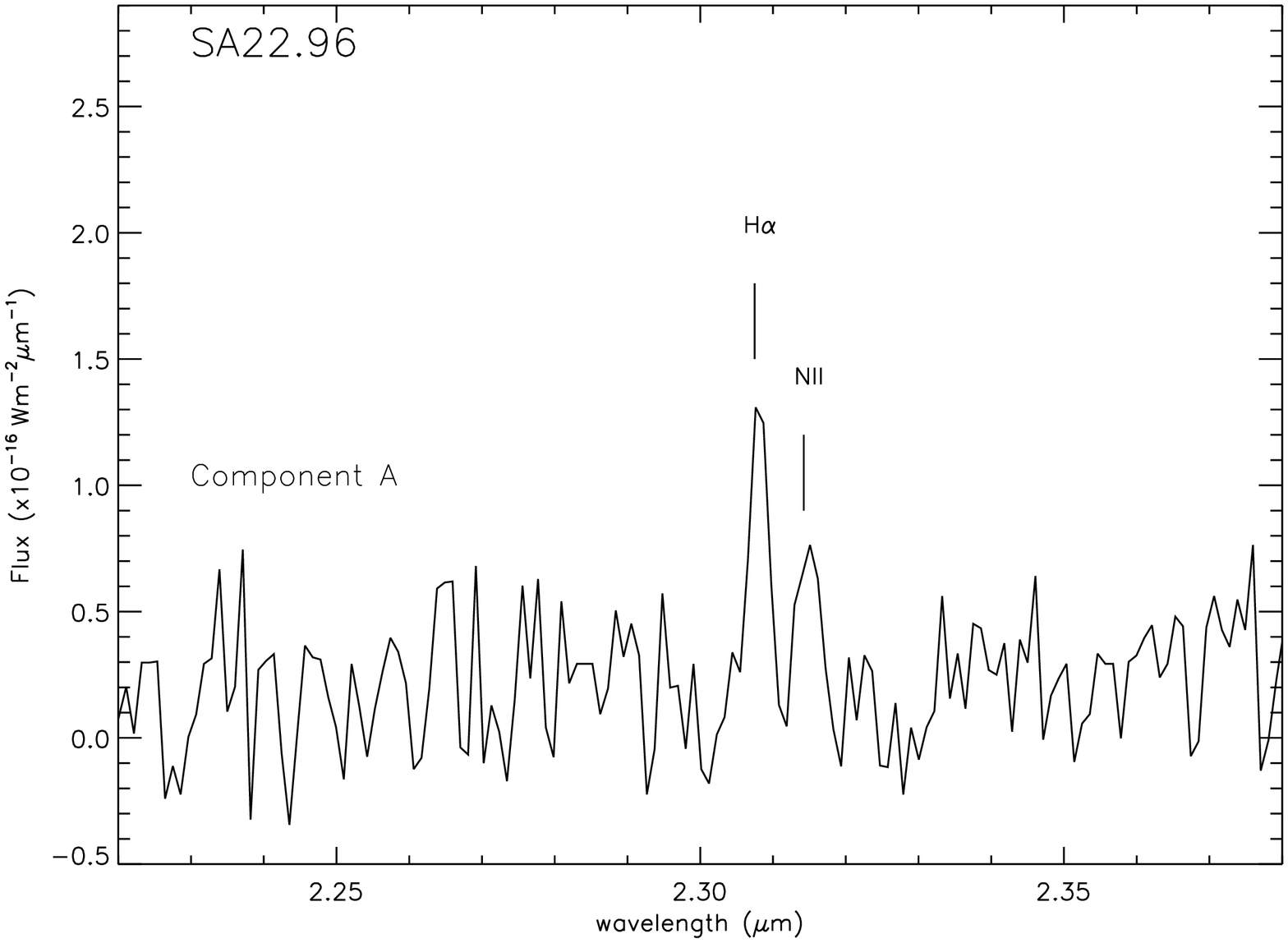,width=2.8in,angle=90}}
\end{minipage}
\begin{minipage}{7.5in}
{\addtolength{\baselineskip}{-3pt}
{\scriptsize 
  \noindent {\bf Figure~6.}  {\it Left:} $HST$ $I_{814}$-band image of
  SA\,22.96 with the H$\alpha$ emission from the UIST IFU observations
  overlaid as contours.  %The continuum imaging shows at least two
%  bright knots seperated by $\sim$2$''$ ($\sim$16\,kpc) in projection,
%  with the brightest H$\alpha$ emission coming from the Western-most
%  component.  
  {\it Right:} Collapsed spectra around the H$\alpha$ emission from the
  Western component ($A$) from the UIST IFU observations of SSA22.96.
  We identify $A$ as the submm source at redshift $z$=2.517 through
  identification of strong H$\alpha$ and [N{\sc ii}] emission at the
  same redshift as derived from the rest-frame UV spectroscopy in
  Chapman et al.\ (2005).  We note that there is tentative evidence for
  weak H$\alpha$ emission from the structure $\sim1.5''$ to the East,
  suggesting it may be associated with the system.

}}
\end{minipage}
\vspace{0.6cm}

\end{figure*}

\section{Observations and Reduction}
\label{sec:obs_red}

\begin{table*}
\begin{center}
{\footnotesize
{\centerline{\sc Table 1.}}
{\centerline{\sc Log of IFU Observations}}
\begin{tabular}{lccccccc}
\hline
\noalign{\smallskip}
SMG          & RA          & Dec       & $z$       & $t_{exp} (ks)$ \\
\hline
\hspace{-0.3cm} UIST/UKIRT\\
SMM\,J163639.01+405635.9 (N2\,850.7)   & 16:36:39.40 & +40:56:38.0 & 1.4880 & 21.6 \\ 
SMM\,J163655.80+405914.0 (N2\,1200.18) & 16:36:55.90 & +40:59:12.0 & 2.5918 & 21.6 \\ 
RG\,J163655.05+410432.0  (N2\,1200.90) & 16:36:55.05 & +41:04:32.0 & 2.1981 & 14.4 \\ 
MIPS\,J142824.0+352619.0               & 14:28:24.40 & +35:26:21.8 & 1.328  & 21.6 \\ 
SMM\,J221804.42+002154.4 (SSA\,22.96)  & 22:18:04.43 & +00:21:53.2 & 2.517  & 10.8 \\ 
\hline
\hspace{-0.3cm} Gemini/GNIRS\\
SMM\,J030227.73+000653.5 (CFRS\,03.15) & 03:02:27.6 & +00:06:52.5  & 1.4076 & 7.2 \\ 
\hline
\hline
\label{table:obs_details}
\end{tabular}
}
\caption{Note: The submm names/position give the 850$\mu$m centroid
whilst the RA and DEC marked denote the 1.4GHz radio centroid.}
\end{center}
\end{table*}

%All but one of
The SMGs in this paper come from surveys at 850 or 1200$\mu$m by
\citet{Scott02,Greve04} and \citet{Webb03}.  These sources were
spectroscopically identified by \citet{Chapman03a,Chapman05a} and
confirmed with precise redshifts from \citet{Swinbank04} based on
accurate radio positions \cite[e.g.][]{Ivison02}.  We also include
the recently discovered hyper-luminous far-infrared-selected galaxy
\b59 from \citet{Borys06} aswell as an optically faint radio galaxy
(OFRG) which was tentatively detected as a millimeter continuum source
in \citet{Greve05} for which a spectroscopic redshift was measured by
Chapman et al.\ (2006).  The coordinates and redshifts for each source
are reported in Table~1 and we briefly review the properties of each
individual source here.

\subsection{Sample}

{\bf CFRS\,03.15:} First identified in the Canada-UK Deep Submm Survey
by \citet{Webb03b}, this submm galaxy has $S_{\rm 850}=4.4\pm1.1$\,mJy
and a 1.4GHz flux density of $S_{\rm 1.4GHz}=226\pm12\mu$Jy.
Spectroscopic observations from \citet{Chapman05a} and
\citet{Swinbank04} yield a redshift of $z=1.4076$ which implies a
luminosity of L$_{FIR}$=5.8$\pm$1.5$\times$10$^{12}\Lsol$.  The
rest-frame UV spectroscopy from \citet{Chapman05a} shows strong C\,{\sc
  iv} emission, which along with the high [N{\sc ii}]/H$\alpha$ line
ratio and apparent underlying broad H$\alpha$ line in the near-infrared
spectra from \citet{Swinbank04} suggests substantial AGN activity.

{\bf N2\,850.7:} This submm-selected galaxy has
$S_{850}$=9.0$\pm$2.4\,mJy, lies at $z=1.488$, and has a far-infrared
luminosity of L$_{\rm FIR}=6.4\pm1.7\times10^{12}\Lsol$
\citep{Chapman05a}.  The existing optical and near-infrared
spectroscopy of this source suggest that the bolometric luminosity is
predominantly arising from a star-burst, with low [N{\sc ii}]/H$\alpha$
and C\,{\sc iv}/Ly$\alpha$ emission line ratios
\citep{Chapman05a,Swinbank04}.  Interferometric observations of the
molecular CO emission in this system by \citet{Greve05} failed to
detect the CO(2-1) line.  If the gas reservoir within this galaxy is
close to the systemic redshift, indicated by the H$\alpha$ emission
line, then this non-detection implies an upper limit of $<1.8\times
10^{10}\Msol$ on the H$_{2}$ gas mass.

{\bf N2\,1200.18:} The near-infrared spectroscopy of N2\,1200.18
indicates a redshift of $z=2.592$ \citep{Swinbank04}, and its
1200$\mu$m flux density of 2.2$\pm$0.6\,$\mu$Jy then corresponds to a
far-infrared luminosity of L$_{\rm FIR}=10.9\pm3.0\times10^{12}\Lsol$
for a reasonable template SED.  The radio flux density $(S_{\rm
  1.4GHz}=\rm 180\pm22\mu Jy)$ is relatively strong, suggesting a
possible contribution from an AGN.  This is confirmed by the existing
longslit optical and near-infrared spectroscopy which shows high [N{\sc
  ii}]/H$\alpha$ and C\,{\sc iv}/Ly$\alpha$ emission line ratios as
well as a broad component underlying the H$\alpha$ emission line (with
FWHM of $2900\pm400\kms$) from a partially obscured AGN
\citep{Chapman05a,Swinbank04}.

{\bf N2\,1200.90:} The 1.4-GHz map of the ELAIS-N2 field shows a
slightly resolved source around this galaxy, with a peak flux density
of $\rm 47\pm10\mu$Jy\,beam$^{-1}$ and a total flux density of $\rm
63\pm21\mu$Jy (Biggs \& Ivison 2006).  A spectroscopic redshift
$z=2.195$ was measured through the identification of Ly$\alpha$ and
C\,{\sc iv} emission lines, the ratio of these suggests weak AGN
activity is possible (Chapman et al.\ 2006).

{\bf \b59:} Far-infrared imaging with the {\it Spitzer} Space Telescope
recently uncovered an apparently hyper-luminous galaxy at $z\sim1.325$
\citep{Borys06}.  Multi-wavelength follow-up of this galaxy at submm
wavelengths suggest a far-infrared luminosity of
3.2$\pm$0.7$\times$10$^{13}\Lsol$.  Unlike most other $z>1$ sources of
comparable luminosity, \b59 lacks any trace of AGN activity, and
therefore the immense luminosity appears to arise due to a
$\gs$\,5000\,$\Msolyr$ star-burst.  However, spectroscopic observations
in the $z$-band have shown that this galaxy is aligned with a
foreground $z$=1.03 elliptical galaxy, suggesting the possibility of
gravitational amplification of the background source.  This would
provide a much more mundane explanation for the apparently immense
luminosity of this galaxy.

{\bf SSA\,22.96:} Rest-frame UV spectroscopy of SSA\,22.96 from
\citet{Chapman05a} indicates a redshift of $z=2.517$ from UV absorption
features, with little indication of AGN activity.  This radio
pre-selected galaxy has 850\,$\mu$m and 1.4\,GHz flux densities of
S$_{850}$=9.0$\pm$2.3mJy and S$_{1.4}$=43.8$\pm$10.1$\mu$Jy which
suggests a far-infrared luminosity of
$L_{FIR}\sim$8$\times$10$^{12}\Lsol$.  %The high resolution
%1.4-GHz radio observations of this field shows an extended source
%aligned along the rest-frame UV slit position \cite{Chapman04a}.

Deconvolving the exact contributions from the star formation and AGN
activity within these galaxies is difficult using just the UV slit
spectroscopy from \citet{Chapman05a}.  Ideally we need to use the
well-developed spectral indicators based on rest-frame optical emission
lines \citep{Veilleux87} which fall in the near-infrared for these
high-redshift galaxies \citep{Swinbank04}.  Going one step further, and
coupling the spatial coverage from an IFU with coverage around the
rest-frame optical emission lines allows us to locate and isolate the
components hosting the AGN in these systems, determine dynamical
masses, as well as searching for extended halos and/or companions.

\subsection{UIST Near-Infrared Integral Field Spectroscopy}

Spectro-imaging observations of five targets were made with the UIST
near-infrared integral field spectrograph on UKIRT.  The UIST IFU uses
an image slicer to take a 3.3$''\times$6.0$''$ field and divides it
into 14 slices of width 0.24$''$. The dispersed spectra from the slices
are reformatted on the detector to provide 2-D spectro-imaging, in our
case using the $HK$ grism, at a spectral resolution of
$\lambda/\Delta\lambda\sim1000$ and covering a wavelength range of
1.4--2.4\,$\mu$m. N2\,850.7 and N2\,1200.18 we observed between 2004
August 20 and September 04, whilst the remaining targets were observed
between 2005 June 25 and June 30\footnotemark.  All observations we
taken $<0.6''$ seeing and photometric conditions.  Observations were
carried out in the standard ABBA configuration in which we chopped by
12$''$ to blank sky to achieve good sky subtraction.  Individual
exposures were 240\,s and each observing block was 7.2\,ks which was
typically repeated two or three times, thus the total integration time
for each object were 10--20\,ks (Table~1).

\footnotetext{Programme ID: U/04A/162 \& U/05A/45: The United Kingdom
  Infrared Telescope is operated by the Joint Astronomy Center on
  behalf on the UK Particle Physics and Astronomy Research Council.}

To reduce the data we used the relevant {\sc orac-dr} pipeline
\citep{Cavanagh03} which sky-subtracts, extracts, wavelength
calibrates, flat-fields, flux calibrates and forms the datacube.  To
accurately align and mosaic the 
individual datacubes together we created white light
(wavelength collapsed) images around the redshifted H$\alpha$ emission
line from each observing block and used the peak intensity to centroid
the object in the IFU datacube.  We then spatially aligned and co-added
the individual data-cubes to create the final mosaic.

\subsection{GNIRS Integral Field Spectroscopy}
We used the GNIRS IFU spectrograph to observe the $z=1.4076$
submm galaxy CFRS\,03.15\footnotemark.  The GNIRS IFU uses an image
slicer to take a 3.2$''\times$4.8$''$ field and divides it into 21
slices of width 0.15$''$.  By re-arranging the slices end-to-end at the
detector, the IFU allows images to be dispersed into full length
spectra whilst preserving the complete 2-D spatial information content.
This allows 3-D spectroscopy across a contiguous 3.2$''\times$4.8$''$
field at 0.15$''$ per pixel.  We used the $H$-band filter with the
32\,lines/mm grating which results in a spectral resolution of
$\lambda/\Delta\lambda\sim$1700
and records a wavelength coverage of between 1.37 and 1.87$\mu$m.
Observations were taken on 2005 October 20 in photometric conditions,
but moderate ($\sim$1.0$''$) seeing.  We used the same observing
strategy as with UKIRT: observations were carried out in the standard ABBA
configuration in which we chopped away to sky by 12$''$ to achieve good
sky subtraction.  Individual exposures were 600\,s seconds and each of
the ABBA observing blocks was repeated three times to give a total
integration time of 7.2\,ks.  To flux calibrate the data, we observed
the UKIRT standard star FS10.  We used the Gemini {\sc iraf} package to
extract, flat-field and wavelength calibrate the data and used {\sc
  idl} to build and combine the datacubes.

\footnotetext{Programme ID: GN-2005B-Q-60: Based on observations
  obtained at the Gemini Observatory, which is operated by the
  Association of Universities for Research in Astronomy, Inc., under a
  cooperative agreement with the NSF on behalf of the Gemini
  partnership: the National Science Foundation (United States), the
  Particle Physics and Astronomy Research Council (United Kingdom), the
  National Research Council (Canada), CONICYT (Chile), the Australian
  Research Council (Australia), CNPq (Brazil) and CONICET (Argentina)}

\subsection {{\it HST} Optical and Near-Infrared Imaging}

{\it HST} Advanced Camera for Surveys (ACS) observations of N2\,850.7,
N2\,1200.18 were obtained from the {\it HST} public
archive\footnotemark (Program ID \#9761) whilst ACS images of
SSA\,22.96 was observed as part of Program \#10145.  The data consist
of dithered exposures with the F814W filter, taken in {\sc lowsky}
conditions using the default four-point {\sc acs-wfc-dither-box}
configuration. This pattern ensures optimal half-pixel sampling along
both coordinates.  The typical integration time were 4.8\,ks.  We
reduced the data using the latest version of the {\sc multidrizzle}
software \citep{Koekemor02} using the default parameters with {\sc
  pixfrac=1} and {\sc scale=1}.  The resulting images have 0.05$''$
pixels and are free from artifacts.  Further details of these
observations are given in Borys et al.\ (2006b)
(see also \citealt{Almaini05}).

{\it HST} WFPC2 observations of CFRS\,03.15 were also obtained from the
{\it HST} public archive.  The data were taken in the F814W and
F450W filters for total of 1.4\,ks each (Program ID \#05996 and
\#06556).  These observations were reduced using the standard {\sc
  stsdas} package in {\sc iraf}.

\footnotetext{Obtained from the Multimission Archive at the Space
  Telescope Science Institute (MAST).  STScI is operated by the
  Association of Universities for Research in Astronomy, Inc., under
  NASA contract NAS5-26555. Support for MAST for non-{\it HST} data is
  provided by the NASA Office of Space Science via grant NAG5-7584 and
  by other grants and contracts.}

{\it HST} NICMOS observations of N2\,850.7 and CFRS\,03.15 were obtained
in Cycle 12, and the targets were observed using the NIC2 camera in the
F160W filter for a total of 2.3\,ks (Program ID \#9856).  We employed
the standard four point spiral dither pattern, {\sc lowsky} conditions
and used the {\sc multiaccum} read-mode.  Each exposure was corrected
for a pedestal offset, and then mosaiced using the {\sc calnicb} task
in {\sc iraf}.  Unfortunately the observation was affected by the South
Atlantic Anomaly (SAA), and extra processing steps were
required\footnote{For a full description, see \\
  http://www.stsci.edu/hst/nicmos/tools/post\_SAA\_tools.html}.  The
final images appear very flat and have very low cosmic ray
contamination.  The reduction and analysis of these data is described
in more detail in Borys et al.\ (2006b).

\section{Analysis \& Results}
\label{sec:analysis}

\subsection{Velocity Determinations}
To accurately determine the redshift (or velocity offsets between
components) in the spectra, we identify and fit both the continuum
level and emission lines (e.g.\ H$\alpha\lambda$6562.8, and [N{\sc
ii}]$\lambda\lambda$6548.1,6583.0 where appropriate) simultaneously
with a flat continuum plus Gaussian profiles using a $\chi^{2}$ fit and
taking into account the increased noise in regions of strong sky
emission.  The error bars on the emission-line wavelength in each
spectrum are derived by allowing the wavelength of the best-fit
Gaussian profile to vary and allowing the signal to drop by
$\Delta\chi^{2}=1$.  The quoted error-bars are therefore conservative
and reflect the signal-to-noise of the data.  The velocity error is
propagated through to the mass determinations in
\S\ref{sec:discussion}.  However, since the signal-to-noise in some of
our observations is modest we also attempt to further quantify the
error-bars on the velocity offsets by creating artificial emission
lines at the resolution of the UIST spectrograph.  We produce spectra
with emission lines with an intrinsic width of 150\,$\kms$ and add
Gaussian random noise.  We then refit the Gaussian profiles to the
data, measuring the velocity offset betwen the input and output
spectrum.  Using 10$^{4}$ simulations at each signal-to-noise, the rms
velocity error between the measured fit and the input velocity as a
function of signal-to-noise (S/N) is: 140$\kms$ at S/N=3, 100$\kms$ at
S/N=5, 65$\kms$ at S/N=7 and 50$\kms$ at S/N=9.  As expected, these
estimates are comparable to the quoted velocity errors derived using
our $\chi^{2}$ estimate.

Before we discuss what can be learnt from our observations as a whole,
we briefly review the results from each of the six galaxies.

\subsection{CFRS\,03.15}

In Fig.~1 we show a true colour {\it $BIH$}-band {\it HST} image of
CFRS\,03.15.  The galaxy shows a compact source surrounded by a diffuse
halo of material in a ring-like structure.  The colours of the nucleus
($B_{450}-I_{814}$=1.8 $I_{814}-H_{160}$=2.5) are consistent with a
star forming galaxy at $z=1.407$, although the ``ring'' of material
surrounding the nucleus is much redder in $(B-I)\gs3.5$, but bluer in
$(I-H)=1.8\pm0.2$, suggesting that [O{\sc ii}]$\lambda$3727 emission
($\lambda_{\rm obs}$=8973\AA) may be contributing significantly to the
$I$-band flux.  There is also a compact knot (or companion)
approximately 1.5$''$ (12.5\,kpc) to the South-West with similar
colours to the nucleus.

From our datacube we construct a wavelength-collapsed H$\alpha$ image
(collapsed between $-$300 and +300\,$\kms$ from the systemic redshift
of \citealt{Chapman05a}) and overlay this on the NICMOS $H$-band image
(Fig.~1).  The $H$-band and H$\alpha$ morphologies are well matched,
and we confirm that the bright knot located $\sim$1.3$''$
($\sim$11\,kpc) to the South-West is associated with the galaxy (the
extended H$\alpha$ morphology and H$\alpha$ velocity shear was also
tentatively detected in the Keck/NIRSPEC spectrum in \citet{Swinbank04}
in which the longslit was placed across both components).  Since the
observations were taken in $\sim$1$''$ seeing, we cannot discuss the
small scale properties of each component in detail (or confirm the
nature of the low surface brightness halo of diffuse material
surrounding component A), however, these will be discussed in Blain et
al.\ (in prep) who use Laser Guided Adaptive Optics assisted $H$-band
Keck/OSIRIS IFU observations of this system.

By extracting spectra from around the two brightest components we
identify a velocity offset between the nucleus and the knot 
of 90$\pm$20$\kms$.  We note that
the [N{\sc ii}]/H$\alpha$ emission line flux ratios of components A and
B are both high (0.95$\pm$0.05 and 1.05$\pm$0.05 respectively),
suggesting that both components may host active AGN. In order to
further search for signatures of non-thermal activity, we attempt to fit
each of the spectra with an underlying broad-line component, but the
improvement in $\chi^2$ over just fitting [N{\sc ii}]+H$\alpha$+[N{\sc
  ii}] is not significant.

\subsection{N2\,850.7}
We detect H$\alpha$ emission at 1.630$\mu$m (in agreement with the UV
and H$\alpha$ redshift from \citealt{Chapman05a} and
\citealt{Swinbank04}) and construct an H$\alpha$ image of N2\,850.7 by
collapsing the datacube between 1.620 and 1.640$\mu$m.  In Fig.~2 we
show the {\it HST} NICMOS image of N2\,850.7 and overlay on this the
H$\alpha$ image from the IFU observations.  The H$\alpha$ morphology of
the central components is spatially extended and therefore to search
for velocity sub-structure, we extract spectra from the regions in the
IFU datacube around the brightest components seen in the {\it HST}
imaging.  From each individual component, the signal-to-noise of the
H$\alpha$ emission is only modest; nevertheless, it is sufficient to
centroid the H$\alpha$ emission line in all cases and so search for
velocity offsets between components.  We identify a velocity offset
between components A and B of $215\pm80\kms$ and an offset between A
and C of $135\pm90\kms$.  The {\it HST} imaging and velocity offsets
suggest that at least two galaxies are undergoing an interaction,
although whether component C belongs to the same structure as B (or A)
will have to wait for higher resolution (adaptive optics) IFU
observations.  We note that the continuum colours of A, B and C are
very similar \citep{Smail04}.

Using the H$\alpha$ luminosity as a star-formation rate indicator, we
derive star-formation rates of 90$\pm$20, 70$\pm$20 and 60$\pm$20
\,$\Msolyr$ from components A, B and C respectively
\citep{Kennicutt98}.  In comparison, the star-formation rate for the
whole system from the far-infrared luminosity is
$1100\pm300$\,$\Msolyr$, implying $\sim 2$ magnitudes of extinction
\citep[q.v.][]{Smail04}.  We reiterate that neither the emission line
flux ratios from the IFU observations nor the rest-frame UV
spectroscopy from \citet{Chapman05a} show any signs of AGN signatures.

\subsection{N2\,1200.18}

The {\it HST} $I_{814}$-band ACS image of N2\,1200.18 ($z=2.5918$)
appears complex with a bright knot and a low-surface-brightness
extension or companion approximately 0.7$''$ (6\,kpc) to the North.
Since the IFU observations of this galaxy cover H$\beta$, [O{\sc
iii}]$\lambda$5007 and H$\alpha$, we construct wavelength-collapsed
images around the [O{\sc iii}]$\lambda$5007 and H$\alpha$ emission
lines and overlay these on the NICMOS $H_{160}$-band image in Fig.~3.
The strongest [O{\sc iii}]$\lambda$5007 emission appears to come from
an unresolved point source in the central, bright $I_{814}$-band
continuum component.  In contrast, the H$\alpha$ is much more extended
and has features which match the extended Northern component seen in
the $I_{814}$-band.  This suggests that the compact continuum source
may host an AGN and indeed, both the strong, spatially unresolved
[O{\sc iii}]$\lambda$5007 and high [O{\sc
iii}]$\lambda$5007/H$\alpha$(=2.0$\pm$0.5) and [N{\sc
ii}]/H$\alpha$(=0.7$\pm$0.1) emission line ratios suggest AGN activity.
We note that the H$\alpha$/H$\beta$ emission line flux ratio of the
central and northern components are 8$\pm$3 and $>$5 respectively.

We can use the emission line flux ratios of [O{\sc iii}]/H$\beta$ and
[N{\sc ii}]/H$\alpha$ of N2\,1200.18 to classify the two components
seen in the UIST IFU data.  The northern component has an [N{\sc
  ii}]/H$\alpha$ emission line flux ratio of 0.6$\pm$0.1, and an [O{\sc
  iii}]$\lambda$5007/H$\beta$ ratio of 0.5$\pm$0.2, together these
place this component it in the LINER region of the diagnostic diagrams
of \citet{Baldwin81}.  The high [O{\sc iii}]$\lambda$5007/H$\alpha$ and
[N{\sc ii}]/H$\alpha$ emission line flux ratios from the central unresolved
component, indicate a Seyfert-2 type AGN.

Using the velocity derived from the H$\alpha$ emission, we find that
the unresolved central component is redshifted from the northern extended
component with a velocity offset of $250\pm100\kms$ across 0.8$''$
($\sim$\,7\,kpc) in projection.

Using deep {\it Chandra} observations of the ELAIS N2 field it is also
possible to compare the X-ray and spectral properties of N2\,1200.18.
This galaxy is detected in the hard (2--8\,keV) X-ray image with a flux
of $14.7\pm2.1\times 10^{-15}$\,erg\,s$^{-1}$\,cm$^{-2}$
\citep{Manners03}.  We convert the observed 2--8\,keV flux to a rest
frame 2--10\,keV luminosity using the equations from
\citet{Alexander03b} assuming a spectral index $\Gamma$=2.  We derive
$L_{\rm X}$=9.1$\pm$1.3$\times$10$^{44}$erg\,s$^{-1}$.  Under the
assumption that the [O{\sc iii}]$\lambda$5007 emission line and the
hard X-ray fluxes are isotropic, they can be used to investigate the
intrinsic power of Seyfert galaxies
\citep{Mulchaey94,Alonso-Herrero97}.  N2\,1200.18 sits comfortably in
the scatter of the L$_{[OIII]}$--L$_{X}$ plot for local Seyferts from
\citet{Mulchaey94}, although it lies at the high luminosity end,
furthering the suggestion that N2\,1200.18 hosts a high luminosity
Seyfert 2 AGN, and also suggesting that increased AGN activity results
in increased photo-ionization of the surrounding material.  This
source, along with low-resolution spectroscopy of the [O{\sc iii}]5007
emission from a larger sample of submm galaxies, is discussed in detail
by Takata et al.\ (2006).

\subsection{N2\,1200.90}

The {\it HST} I$_{814}$ image of this galaxy shows a compact morphology
($\ls$0.2$''$ (1.6\,kpc) FWHM), which is matched by the H$\alpha$
morphology from the UIST IFU observations.  The H$\alpha$ emission from
this galaxy is spatially unresolved and we detect no significant
velocity structure across the galaxy.  The line width of the H$\alpha$
emission is 420$\pm$100\,km\,s$^{-1}$ and the integrated line flux of
0.5$\times$10$^{-18}$\,Wm$^{-2}$ then suggests a SFR of
150$\pm$50\,$\Msolyr$ \citep{Kennicutt98}.

\subsection{\b59}

The discovery that there is a fortuitous alignment of a massive
foreground elliptical galaxy at $z$=1.03, superimposed onto the
$z$=1.325 far-infrared-luminous 
galaxy \citep{Borys06} suggests
the possibility that \b59 may be gravitationally lensed.  To this end,
we targeted the H$\alpha$ emission with the UIST IFU in order to search
for morphological
signatures of strong lensing which might
explain the apparent immense luminosity of \b59.

In Fig.~5 we show the continuum subtracted H$\alpha$ emission line map
of \b59 overlaid on the $K$-band continuum image generated from the
IFU.  The H$\alpha$ emission is spatially extended, across $\sim$1$''$
in projection, however, the $H$ and $K$-band light is much more widely
distributed, showing likely contamination by light from the forground
lens.  The H$\alpha$ emission appears elongated, but with an alignment
which is not naturally ascribed to gravitational lensing (we would
expect the brightest images to straddle the long-axis of the foreground
lens potential). To search for velocity structure, we extract a series
of independent spectra from the datacube along the arc, and place a
limit of $\ls$\,125\,$\kms$ on any possible velocity shear.  We show
the collapsed, one-dimensional spectrum around the H$\alpha$ emission
in Fig.~5.  The lack of velocity structure may be a further indication
that the H$\alpha$ morphology arises from two merging images of the
same background source, with the third (much lower surface brightness)
image undetectable to the South West.  Whilst such a lens model is
somewhat contrived, it would potentially yield an amplification $\gg$10
for the background source.

\subsection{SSA\,22.96}

The {\it HST} ACS $I$-band imaging of SSA\,22.96 shows several components
around the submm galaxy (labelled A).  Component A is the radio
counterpart in the 1.4-GHz map from \cite{Chapman04b}.  To align the
IFU datacube with the optical imaging we construct a H$\alpha$ image of
SSA\,22.96 by collapsing the datacube between 2.29 and 2.33$\mu$m and
overlay this on the {\it HST} $I_{814}$-band image in Fig~6.  We extract a
spectrum from the datacube around the position of the submm galaxy and
also show this in Fig.~6.  The H$\alpha$ redshift of component A is
in excellent agreement with the UV redshift from \citet{Chapman05a}.

The rest-frame UV spectroscopy from \citet{Chapman05a} is measured from
absorption lines, yet the high [N{\sc ii}]/H$\alpha$ emission line flux
ratio (0.6$\pm$0.2) we see may be more consistent with either high
metallicity or AGN activity \citep{Veilleux87}.  We note that this
galaxy is not detected in the archival 0.4--10\,KeV XMM X-ray image 
of this field to a
flux limit of 6$\times$10$^{-15}$erg\,s$^{-1}$\,cm$^{-2}$ (which
corresponds to L$_{X}<$3$\times$10$^{44}$erg\,s$^{-1}$).

\section{Discussion}
\label{sec:discussion}

In this paper, we have spatially mapped the rest-frame optical line
emission in a sample of six high redshift, luminous submm galaxies.  We
also include in our discussion two similar galaxies which also have
published IFU observations (N2\,850.4 and SMMJ\,14011+0252;
\citealt{Swinbank04,Tecza04}). The most striking feature in the
combined sample of eight submm galaxies is the prevelance of
galactic-scale multiple components: five of the eight appear to
comprise two or more galactic-scale extensions/components on 1--2$''$
(8--16\,kpc) scales.  Our IFU observations also show that velocity
structures can be identified within individual components (e.g.\ on
$\ls$0.5$''$, 4\,kpc scales).

Such velocity structures are not only observed through the nebular
emission: recently \citet{Greve05} and \citet{Tacconi06} have published
mm-wavelength observations of twelve submm galaxies.  Several of these
systems show double-peaked CO emission line spectra, with velocity
offsets of up to 800$\kms$ and line widths of up to 1000$\kms$ (FWHM).
The individual CO peaks often have similar intensities
($I_{CO_{1}}$/$I_{CO_{2}}$=0.7$\pm$0.1 with a range of
$I_{CO_{1}}/I_{CO_{2}}$=0.3--0.9) suggesting that many of the submm
galaxies could contain massive gas disks \citep{Genzel03}.  This
interpretation is supported by examples of ULIRGS with gas rich disks
in the local Universe (e.g.\ NGC\,6240; \citealt{Tacconi99}).  However,
the obvious interacting/merger nature of many of these distant submm
galaxies from our {\it HST} and IFU observations strongly suggests that
the velocity offsets we measure arise from merging systems.  The IFU
observations also show that the components have large differences in
their energetics and AGN contributions over 2--20\,kpc scales.  %This
interpretation is supported by recent observations of
%local ($z\ls 0.1$) ULIRGs which have shown that galaxies with
%$L\gs$\,10$^{12}$L$_{\odot}$ typically comprise systems two (or more)
%interacting components with mass ratios 1:1 to 1:3 \citep{Dasyra06}.
%Whilst there are only eight galaxies with spatially resolved
%spectroscopy in our sample, it is also interesting to note that two of
%these appear to comprise at least three components (N2 850.4 and
%N2\,850.7), the others comprising binary interactions.  In comparison,
%85\% of local ULIRGs are composed of two or more interacting systems
%\citep{Bushouse02} and therefore our observations provide complementary
%evidence that the multi-component nature of local ULIRGs in mirrored in
%high redshift submillimetre galaxies.

\subsection{Mass Estimates}
\label{sec:MassEstimates}

Constraining the stellar, gas and dynamical masses of submm galaxies is
crucial if we are to understand their evolution.  Recent dynamical
masses for SMGs from CO spectroscopy with millimetre interferometers
indicate a dynamical mass of $\log($M$_{dyn})=11.1\pm0.3$ and gas
masses of $\log($M$_{gas})=10.5\pm0.2$ within the central
$\sim$\,10\,kpc \citep{Greve05,Tacconi06}. In addition, estimates of
the stellar masses for submm galaxies have been given by
\citet{Borys05}.  These are derived by fitting simple stellar
population models to the spectral energy distributions of submm
galaxies across the restframe ultraviolet, optical and near-infrared
wavebands.  The inclusion of restframe near-infrared data into these
fits improves their insensitivity to dust extinction and thus provides
much more reliable measures of the luminosity of the stellar
populations in these systems. In this way, \citet{Borys05} derived a
median stellar mass of $\log($M$_{\ast})=11.4\pm0.4$ for a small sample
of spectroscopically identified submm galaxies in the GOODS-N region.

We are also able to place limits on their dynamical masses in this
population using the projected velocity offsets and spatial separations
in our IFU observations.  In Fig.~\ref{fig:dVdR} we construct a
velocity offset versus spatial offset diagram for all submm galaxies
which show multi-components in their spatially resolved spectra, either
from IFU spectroscopy (this work; \citealt{Swinbank05b,Tecza04});
resolved CO spectroscopy \citep{Greve05,Tacconi06}; or longslit
observations \citep{Swinbank04}.  We note that no two data-points in
Fig.~\ref{fig:dVdR} represent the same sub-mm galaxy: the only
multi-component SMG which has both CO and IFU observations (N2\,850.4)
is represented by an IFU data-point in the figure (the measured
kinematics in N2\,850.4 from IFU and CO observations give consistent
results and are discussed in \citet{Swinbank05b}; see also
\citet{Tacconi06}).

To investigate the masses of these systems we assume that the velocity
offsets between the different components eventually transform into
motions in a single, relaxed potential well, i.e.\ they reflect virial
motions.  We therefore overlay the expected projected velocity
dispersion as a function of radius for an elliptical galaxy with an NFW
profile \citep{NFW,Lokas01}.  We assume a concentration of $c=7$ and a
virial radius of 200\,kpc -- typical of a massive elliptical galaxy
\citep{Mamon05,romanowsky03}.  Interpreting the velocity offsets in
this way, we derive a median dynamical mass of
5$\pm$3$\times$10$^{11}\Msol$ within the virial radius for these
galaxies.

Before we compare the dynamical mass estimates with the stellar and gas
masses, we note first that there are clearly uncertainties in applying
the virial theorem and attempting to measure the mass of a
(non-stationary) merging systems.  The following caveats must be
considered:

1) Whilst these are merging systems and out of equilibrium, the virial
theorem may still be applied if an extra term is included to reflect
the moment of inertia of the system (d$^{2}$I/dt$^{2}$): such that
2T+W=d$^{2}$I/dt$^{2}$.  Since the moment of inertia is strongly
weighted towards large radii (I$\propto\int r^{2}$dm), relatively minor
parts of the galaxy (such as tidal tails) can have a large impact on
$I$ and hence the virial mass estimate.  However, in many of these
systems, the velocity dispersions (and star-formation activity) of the
individual components (and the fact that two SMGs appear to comprise
double AGN) suggest that we are not measuring velocities for ``minor''
components, but rather these are likely to be the dominant components
of the merger.

2) The virial theorem is strictly only valid if the components we
observe follow the distribution of mass in the galaxy.  Since the
components we measure are the likely nuclei of the progenitor galaxies
they will be more concentrated (due to dynamical friction) than the
overall mass distribution and may produce an underestimate of the
dynamical mass.

3) The virial theorem should only be applied if the merging components
have similar mass ratios.  The velocity dispersions (and star formation
activity) of the individual components we measure are often comparable
arguing that the mass ratios of the merging components are similar.  It
is also useful to note that recent observations of local ($z\ls 0.1$)
ULIRGs which have shown that galaxies with $L\gs$\,10$^{12}\Lsol$
typically comprise systems two (or more) interacting components with
mass ratios 1:1 to 1:3 \citep{Dasyra06}.

To gauge the validity of the mass estimate from the virial theorem, we
can make a further estimate of the dynamical masses: if we take the
velocity offsets and assume these are point masses (i.e.\ assume
$M=\Delta v^{2}/\Delta rG$), we derive masses of
2.5$\pm$1.5$\times$10$^{11}\Msol$.  We view this estimate as more
uncertain than that using the virial estimator since this estimate
requires placing all of the mass of the progenitors into point masses
on circular orbits.  The estimate based on the point-mass model is
within the range we quote, and so we consider
5$\pm$3$\times$10$^{11}\Msol$ a reasonable estimate of the mass and its
uncertainty in the typical submm galaxy in our sample.  However, we
note that these estimates would benefit considerably from (high angular
resolution) adaptive optics IFU observations coupled with a theoretical
study of the distribution of detectable emission line gas within dusty,
merging systems at high redshifts.

\subsubsection{Comparison with Stellar Mass Estimates}

\citet{Borys05} estimated restframe 2.2-$\mu$m luminosities for a
sample of spectroscopically-confirmed submillimetre galaxies in the
GOODS-N region from broad-band photometry covering $UBVRIzJK$+IRAC
3.6/4.5/5.8/8$\mu$m -- which spans the restframe UV, optical and
near-infrared at the redshifts of their sample, $z=0.56$--2.91.  They
find $M_{K}=-26.8\pm0.4$ from their sample of 10 galaxies at $z>1.5$.
They then adopted a typical light-to-mass ratio for the stellar
population of L$_K$/M=3.2, based on an estimated mean age of $\sim
250$\,Myrs from fitting simple stellar population models to the
broad-band SEDs, and so derive typical stellar masses of
$\log($M$_{\ast})=11.4\pm0.4$ \citep{Borys05}.  This is a significant
stellar mass for a galaxy at $z\sim 2$ and so \citet{Borys05} discuss
in detail the potential uncertainties in their estimates, in particular
whether it is possible to reduce the stellar masses.  The most obvious
way to achieve this would be to adopt a higher L$_K$/M ratio,
equivalent to assuming a younger age for the stellar population.
However, to achieve an order-of-magnitude reduction, i.e.\ a
light-to-mass ratio of L$_K$/M\,$\sim30$, would require a typical age
of $\sim 10$\,Myrs \citep{Leitherer99}.  Given the observed space
density of submillimetre galaxies, such very young ages and hence
high-duty cycles imply very large volume densities for the progenitor
population ($\sim 10^{-3}$\,Mpc$^{-3}$) which is at least an order of
magnitude greater than the estimates of the masses and clustering of
this population \citep{Blain04a,Chapman05a}.

It is also possible that AGN contribute to the measured 2.2-$\mu$m
luminosities, leading to an over-estimation of the stellar masses. As
\citet{Borys05} point out, the shape of the spectral energy
distributions for typical submillimetre galaxies is well-described by a
stellar model and indeed they remove from their analysis the small
number of galaxies for which an AGN is potentially dominant in the
restframe near-infrared.  Nevertheless, we can take an extreme view
point and ask if AGN produce all of the luminosity in the reddest bands
in these sources, how much would that influence the estimated stellar
luminosity?  We use the average near-infrared spectral index for QSOs
from \citet{Simpson00}, $\alpha = 0.98$, assume that all of the
observed 8.0-$\mu$m luminosity is arising from the AGN and estimate its
contribution to the luminosity at a restframe wavelength of $\sim
1.6\mu$m (where the stellar contribution should peak).  Using this set
of extreme assumptions, we find a median reduction in the stellar
luminosity of only $4\times$, suggesting that a more realistic estimate
taking into account the stellar contribution at 8.0$\mu$m, is likely to
yield a change of $\ls 2\times$, within the error range quoted by
\citet{Borys05}.  It is also worth noting at this point that
\citet{Alexander05a} have shown that while AGN are ubiquitous in SMGs,
their typical contribution to the bolometric emission is $\lsim 10$\%
suggesting that any contamination to the 2.2$\mu$m flux from AGN
activity is minimal and unlikely to bias our result.

Finally, we note that recent stellar population synthesis modelling by
Maraston et al. (2006) has shown that thermally pulsating asymptotic
giant branch (TP-AGB) stars in post starburst galaxies can contribute
significantly to the rest-frame K-band luminosity on
$\gsim200$--1000Myr timescales.  Indeed, the predicted stellar
light-to-mass ratios at 250\,Myr is L$_K$/M$\sim6$--8
(\citealt{Maraston98,Maraston06}) and therefore may act to reduce our
estimated stellar mass by a factor of up to 3$\times$ if the luminosity
weighted stellar populations in the SMGs are dominated by stars with
ages of $\gsim200$--1000Myr.  However, in the case of an on-going
starburst (as is relevant for SMGs) the importance of TP-AGB phase is
likely to be lessened and we estimate a maximum of L$_K$/M\,$\sim$\,5
at 450\,Myr and more typically L$_K$/M\,$\lsim$\,4 for the likely ages
of SMGs, suggesting that this phase is unlikely to bias our results
significantly.

Hence, unless there is a conspiracy between a number of factors, it
appears unlikely that the estimated stellar masses for the
submillimetre population are over-estimated by more than a factor of
2--$3\times$.  We conclude therefore that the typical stellar and
dynamical masses for submillimetre galaxies are likely to be similar in
magnitude.  We estimate a mass ratio of $M_{dyn}/M_{\ast}\sim2$, with a
considerable uncertainty.  This suggests that, as with local luminous
ellipticals, the central regions of these galaxies are
baryon-dominated.

\setcounter{figure}{6}
\begin{figure}
  \centerline{\psfig{file=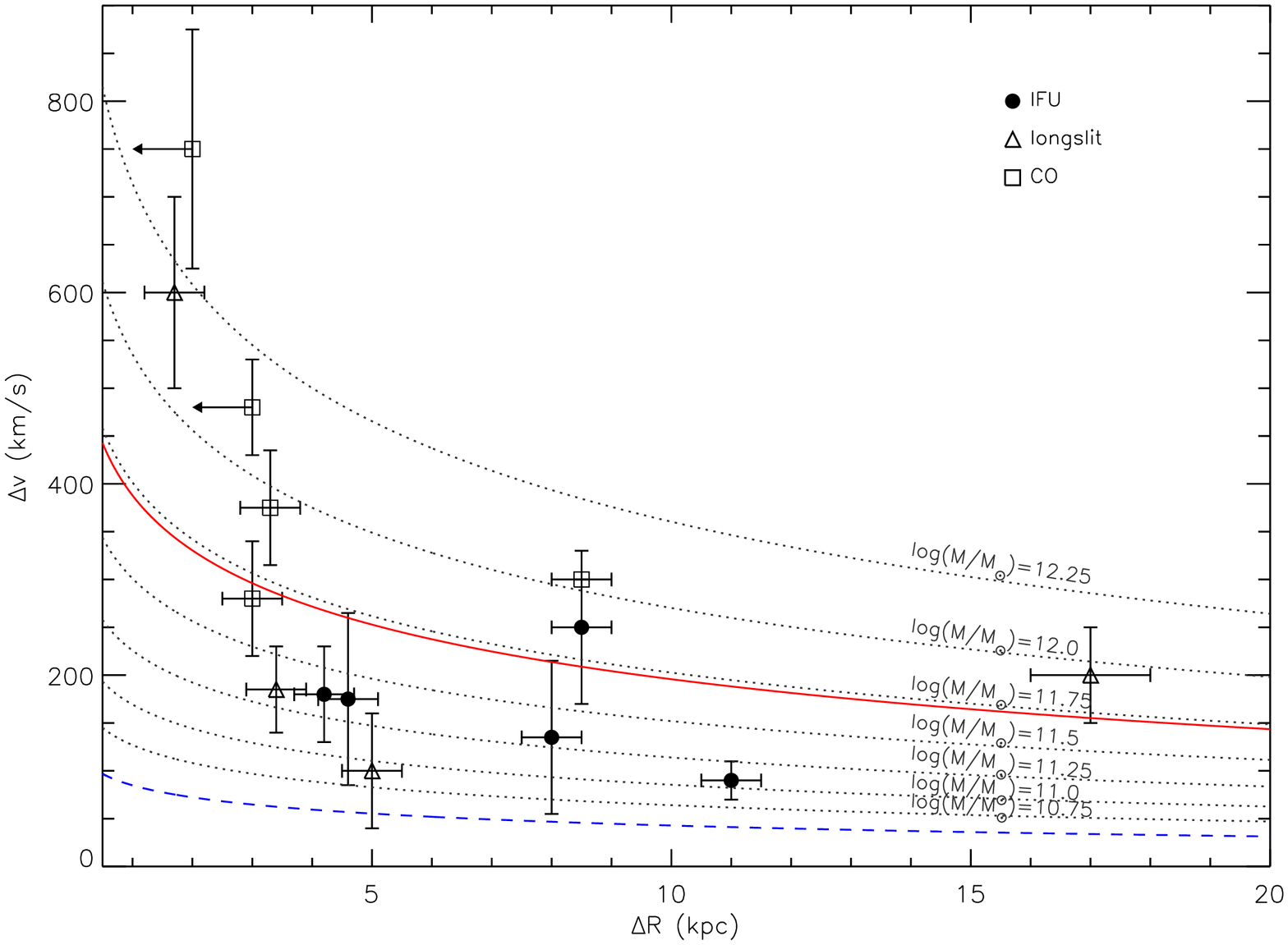,angle=90,width=4in}}
  \caption{Velocity offset versus spatial offset for the multiple
    components within submm galaxies from IFU, longslit or CO
    observations.  The dotted lines show the expected profiles as a
    function of halo mass for a galaxy with an NFW profile and a
    concentration parameter typical of elliptical galaxies. Treating
    the observations as an ensemble, we derive a dynamical mass for the
    SMGs of $M$=5$\pm$3$\times$10$^{11}\Msol$ (shown by the solid
    line).  For comparison, the dashed line shows the profile for the
    minimum stellar mass derived from Borys et al.\ (2005).}
\label{fig:dVdR}
\end{figure}

\subsection{Space Densities and Comparison with Local Populations}

In order to test the connection between SMGs and local galaxy
populations we can evolve their stellar populations to the present day
and determine whether their descendents would be consistent with
scaling relations for local galaxies, such as the Faber-Jackson
relation \citep{Faber76}.  This will test claims that SMGs evolve into
massive ellipticals at the present day \citep[e.g.][]{Lilly99}.  The
velocity offsets we measure in \S\ref{sec:MassEstimates} implies an
average velocity dispersion of approximately $\sigma=240\pm80\kms$.
The Faber-Jackson (FJ) relation for local early-type galaxies indicates
that this dispersion corresponds to an $r$-band magnitude of
$M_{r}=-22.7\pm0.5$ \citep{Bernardi05} or $M_{K}=-25.3\pm0.5$ (assuming
$(r-K)=2.6$, typical for an elliptical galaxy at the present day).  Is
the expected fading of the stellar populations within the SMGs, with
$M_{K}=-26.8\pm0.4$ at $z\sim 2$, sufficient to reconcile them with the
$z=0$ FJ relation?

The key to estimating the degree of fading of the SMGs is to determine
the likely ages of the stellar populations.  Since the starburst
lifetimes in SMGs depends on the mass of the molecular gas reservoir
and the star-formation rate, \citet{Greve05} used CO(3-2) transitions
for 18 SMGs (and assuming a CO--H$_{2}$ conversion factor) to infer gas
masses, obtaining $M(H_2)\sim 3\times10^{10}\Msol$.  Recently, Hainline
et al.\ (2006) have reported CO(1-0) observations of an SMG using the
Green Bank Telescope and derive a H$_{2}$ gas mass four times larger
than that from the CO(4-3) transition of the same source (implying that
$J_{upper}>$3 transitions of CO significantly underestimate the mass of
cold, diffuse molecular gas in SMGs).  This suggests that the typical
gas masses of SMGs are likely to be $M(H_2)\sim 1\times10^{11}\Msol$.
Star formation rates can be derived from the bolometric emission of the
source, assuming this is powered by star formation.  \citet{Greve05}
used 850\,$\mu$m and 1.4\,GHz fluxes and assumed the far-infrared-radio
correlation to determine bolometric luminosities for their sample.
They then assumed a conservative contribution of 50\% to the bolometric
luminosity from a AGN, and so estimated a median star-formation rates
of 700\,$\Msolyr$ for the galaxies in their sample (integrating the
Salpeter IMF between 1 and 120$\Msolyr$).  Two recent results influence
this conclusion, but essentially leave it unchanged.  Firstly,
\citet{Alexander05a} have shown that while AGN are ubiquitous in SMGs,
their typical contribution to the bolometric emission is $\lsim 10$\%,
which would increase the star-formation rate estimated by
\cite{Greve05} by $\sim 2\times$.  However, Kovacs et al.\ (2006) have
used 350\,$\mu$m photometry to improve the estimates of the bolometric
luminosities from the far-infrared SEDs of SMGs and show that the
star-formation rates may be overestimated by a factor $\lsim2\times$,
cancelling this increase.  Hence, our current best-estimate of the gas
consumption timescales in SMGs is: $\tau_{\rm SMG}\sim
M(H_{2})/SFR\sim1\times10^{11}\Msol/700\Msolyr\sim 150$\,Myr, although
this estimate clearly has significant uncertainties.  The reader should
note that a burst with this duration and star-formation rate can
account for all of the observed stellar mass in a typical SMG,
suggesting that the activity we are witnessing is likely to be the
origin of the majority of the stellar mass in these galaxies
\citep{Smail04}.

We can next compare these burst lifetimes with those required in order
for the SMGs to lie on the FJ relation at the present day using stellar
synthesis modelling.  For SMGs with $z>1.5$, the average rest-frame
$K$-band magnitude from \citet{Borys05} is M$_{K}=-26.8\pm0.4$.  We
start by calculating the fading of a simple instantaneous burst to the
present day.  We use {\sc starburst99} \citep{Leitherer99} with a
Salpeter IMF \citep{Salpeter55} and an instantaneous star-burst which
predicts that between 100\,Myr and 10\,Gyr there should be
approximately $\Delta$M$_{K}=3.4$\,mags of fading suggesting the
present day descendants should have M$_{K}\sim-23.4\pm0.4$.  This is
approximately 2 magnitudes fainter than the FJ relation, given the
expected velocity dispersion as found in \S\ref{sec:MassEstimates}.
Similarly, an instantaneous burst using the models of Maraston et al.\ 
(2006) suggests that the fading between 100\,Myr and 10\,Gyr is
approximately $\Delta$M$_{K}=2.1$--2.4 (depending upon the metallicity)
-- which is still $\gsim 0.6$ magnitudes fainter than the FJ
prediction, given our expected velocity dispersion.

However, these instantaneous burst models are unphysical for SMGs which
are still forming stars at the current epoch. So, we derive the fading
from a more realistic model: a constant burst of star-formation of
duration 100, 200 and 300\,Myrs using {\sc pegase} \citep{Fioc04}.  For
a 100\,Myr burst (observed at 50\,Myr, since we expect on average to
see SMGs half way through their active phase) the fading between the
epoch of observation and the present day is $\Delta$M$_{K}=3.2$, $\sim
1.5$\,magnitudes too faint compared to the FJ relation for the velocity
dispersion.  Taking longer burst durations of 200 and 300\,Myrs
(observed at 100 and 150\,Myr respectively) the fading is
$\Delta$M$_{K}=2.7$ and 2.3 magnitudes respectively -- approximately
1.2 and 0.8 magnitudes off the FJ relation.  Thus for our preferred
burst lifetime, 300\,Myrs, observed half way through the burst, the
predicted present-day magnitude is M$_{K}\sim-24.5\pm0.4$ compared to
the $z=0$ value of $M_{K}=-25.3\pm0.5$ from the FJ relation.

These predictions are of course sensitive to the choice of IMF,
changing from a Salpeter to a Kroupa IMF and for a 300-Myr burst the
predicted fading reduces to $\Delta$M$_{K}=1.7$, indicating a
present-day luminosity of M$_{K}\sim-25.1\pm0.4$ and placing the SMGs
within $\sim0.2$ magnitudes of the $z=0$ FJ relation.  We note that
such a change would be at odds with the ``top-heavy'' IMF needed in
semi-analytic models of galaxy formation to reproduce the basic
properties of SMGs (e.g.\, \citealt{Baugh04}; see also
\citealt{Blain99b,Blain99c}).

%These predictions are of course sensitive to the choice of IMF.
%Adopting a Kroupa IMF \citep{Kroupa93} for a 300\,Myr burst
%% the depletion timescale is
%%reduced by a factor 2$\times$ (since the IMF reflects on the
%%star-formation rate).  For a 150-Myr burst 
%the predicted fading reduces to $\Delta$M$_{K}=2.1$, indicating a
%present-day luminosity of M$_{K}\sim-24.7\pm0.4$ and placing the SMGs
%within $\sim0.6$ magnitudes of the $z=0$ FJ relation.  We note that
%such a change would be at odds with the ``top-heavy'' IMF needed in
%semi-analytic models of galaxy formation to reproduce the basic
%properties for SMGs \citep[e.g.][]{Baugh04}).%  We also note that IMF
%% reduces the predicted star-formation rate (between 0.1 and 100$Msol$)
%% by a factor of two, and so reduced the gas depletion timescale to
%% $\sim$100\,Myr.

Hence, whilst there are a number of uncertainties in estimating both
the starburst lifetimes and the fading of SMGs to the present day,
longer burst durations ($\sim 300$\,Myrs) are favoured in order for
SMGs to evolve towards the scaling relations for elliptical galaxies at
the present day.  Burst timescales of this length are also consistent
with the expected time to consume the gas reservoirs in these galaxies
and with the build up of the present stellar mass in the systems.

We can also investigate how the SMG burst lifetimes influence the space
densities of their expected descendants at the present day.  The full
SMG redshift distribution from \citet{Chapman05a} is well fit by a
Gaussian with a mean redshift of $<z>=2.2$ and a with, $\sigma_{z}\sim
1.3$ (accounting for incompleteness in the spectroscopic desert).  Thus
the SMGs populate an epoch of about $\tau_{obs}=1.5$\,Gyr in duration.
Since $\phi$=$\rho_{\rm SMG}\tau_{\rm obs}/\tau_{\rm burst}$ (where
$\phi$ is the comoving space density of the descendant population,
$\tau_{\rm burst}$ is the burst lifetime, and $\rho_{\rm SMG}$ is the
observed space density) an estimate of $\tau_{\rm burst}$ allows us to
infer $\phi$.  We first estimate the observed space density ($\rho_{\rm
  SMG}$) by taking the observed surface density of $>$5\,mJy SMGs from
\citet{Chapman05a} ($\sim$1000 per square degree) which suggests that
between $z$=0.9--3.5 the volume density should be
$\sim2.5\times10^{-5}$Mpc$^{-3}$ (this calculation assumes that 67\% of
the SMGs lie within the 1$\sigma$ limits and includes the
incompleteness correction in both the redshifts and radio-detection
limits).  Using our estimated lifetime of $\tau_{\rm
  burst}\sim300$\,Myr, the corrected space density (assuming a galaxy
only goes through one SMG phase in its lifetime) is therefore $2.5
\times 10^{-5}\times1.5$\,Gyr/300\,Myr\,$\sim
1.3\times10^{-4}$\,Mpc$^{-3}$ for a population with predicted present
day $K$-band absolute magnitude of M$_K\sim -25.1$.  Integrating down
the $z=0$ luminosity function of early-type galaxies from
\citet{Croton05} (see also Wake et al.\ 2006), to a median magnitude of
$M_K\sim-25.1$ we determine a space density of
$9\times10^{-5}$\,Mpc$^{-3}$ (we note that using the LRG luminosity
function from Wake et al.\ 2006 we obtain
$1.1.\times10^{-4}$\,Mpc$^{-3}$).  Thus if SMGs have burst durations of
approximately 300\,Myr, as indicated by their gas consumption
timescales, then their predicted fading would place them on (or
near-to) the present day FJ relation and the observed space density of
this population at $z\sim2$ (corrected for the duty-cycle of the
bursts) would account for the formation of the whole population of
early type galaxies with M${_K}<-25.1$, ($\gsim 3 L^{*}_{K}$) seen at
$z\sim0$.

Further evidence for the evolution of SMGs into massive ellipticals has
also be presented in \citet{Tacconi06} who used high resolution
millimeter imaging to derive matter volume and surface densities of
$\sim$\,100\,cm$^{-3}$ and 5000\,$\Msol$pc$^{-2}$ for the SMGs,
comparable to those in massive local ellipticals and Sa bulges.
Similarly, \citet{Blain04a} provide tentative evidence for clustering
of SMGs at a level consistent with the progenitors of $>L^\ast$
elliptical galaxies.  These provide additional support for our
conclusion that SMGs can be connected through a self-consistent
evolutionary model to the local luminous galaxy population, which
reproduces both the mass--luminosity relationship and luminosity
distributions of local elliptical galaxies.

\section{Conclusions}
\label{sec:conc}

In this paper we have studied the rest-frame optical emission line
structures and dynamics of six high redshift, luminous submm galaxies.
Including two sources from the literature with similar observations, we
find that of the eight submillimetre galaxies with resolved
spectroscopy, at least five appear to comprise two or more dynamical
sub-components.  The average velocity offsets between these components
is $\sim180\kms$ across a projected spatial scale of 8\,kpc.  The
obvious merging/interacting nature of these systems suggests there are
analogous to the multi-component nature of the, typically less
bolometrically luminous and slightly more compact, ULIRGs in the local
Universe.

Since our IFU observations allow us to disentangle AGN and
star-burst-like components (e.g.\ from [N{\sc ii}]/H$\alpha$ flux
ratios), it is perhaps significant that two of the eight submm galaxies
shows possible signs of AGN activity in two spatially-resolved
components on 1--2$''$ (10--20\,kpc) scales.  Deep X-ray surveys of the
{\it Chandra} Deep Field North have produced similar results:
\citet{Alexander03a} report that two of seven submm galaxies in this
region are individually associated with pairs of X-ray sources on
similar spatial scales (approximately one galactic diameter).  Since
the chance of a false association is $\ls1$\%, it may be that we are
witnessing the interaction or merging of AGNs in these sources (a low
redshift example of this binary AGN activity is seen in the ULIRG
NGC\,6240; \citealt{Komossa03}).  The double AGN in some of these
systems, suggests mergers may be responsible for disturbing the gas
orbits in both components, hence fuelling subsequent AGN growth which
currently lags behind a significant pre-existing stellar population
\citep{Alexander03a,Smail03a}.

By combining the spatially-resolved kinematic information from IFU, CO
and longslit observations and assuming the velocity offsets we measure
are eventually transformed into velocity dispersions in a
pressure-supported galaxy, we derive dynamical masses of
5$\pm$3$\times$10$^{11}\Msol$ for typical luminous submillimetre
galaxies.  This is similar to the dynamical masses found using
H$\alpha$ line widths \citep{Swinbank04} and resolved CO spectroscopy
\citep{Greve05,Tacconi06} and a factor of $\sim 2\times$ greater than
the recent stellar mass estimates from \citet{Borys05}, suggesting that
the central regions of these galaxies are baryon dominated.

We combine our dynamical measurement with estimates of the stellar
luminosities for this population from \citet{Borys05}. We use these to
constrain their likely evolution and so compare the properties expected
for the descendents of submillimetre galaxies with $z=0$ galaxy
populations.  We update the gas depletion time scales estimated by
\citet{Greve05} using recent results on the gas masses, bolometric
luminosities and AGN contributions for this population, deriving
typical gas depletion times of $\sim 150$\,Myr.  Adopting a typical
burst duration of twice this age, $\sim 300$\,Myr, we predict the
evolution of their stellar populations and compare these to the
Faber-Jackson relation for local ellipticals.  For a $\sim 300$\,Myr
burst with a Kroupa IMF, we find that the expected fading would result
in the descendents of the SMGs lying within $\sim 0.2$\,mags of the FJ
relation at their measured velocity dispersion.  The same estimate of
the lifetime can be used to correct the apparent space density of SMGs
(assuming a galaxy only goes through one SMG phase in its lifetime) to
derive a space density for the descendents of this population of $\sim
1.3\times10^{-4}$Mpc$^{-3}$ for a population with $z=0$ absolute
luminosities of M$_K\sim -25.1$.
%  This space density is very comparable
%to that of early type galaxies with M$_{K}\lsim -25$ at the present
%day.  
Hence if SMGs have burst durations of $\sim 300$\,Myr, as indicated by
gas consumption timescales, then their predicted fading places them
close to the present-day FJ relation.  Moreover, the observed
population at $z\sim2$ is very comparable to that of luminous
ellipticals with M$_{K}\lsim -25.1$ ($\sim 3 L^{*}_{K}$) at the present
day.
% then sufficient to account for the formation of of $\sim 2 L^{*}_{K}$
% luminous ellipticals seen at $z\sim0$.

Our observations show the power of using integral field spectroscopy
around the rest-frame optical emission lines to probe the dynamics and
power sources of these galaxies on $\gs$5\,kpc scales.  With
forthcoming adaptive optics integral field spectroscopy on 8- and 10-m
telescopes (e.g.\ OSIRIS on Keck, NIFS on Gemini and SINFONI on the
VLT), we will soon be able to probe the dynamics, metallicities and
star-formation properties on $\ls$0.1$''$ (sub-kpc at $z=2$) scales.
Furthermore, the increased light grasp of large aperture telescopes
will allow us to spatially resolve and map the absorption line spectra
of these galaxies (including important diagnostic lines such as Ca{\sc
  ii} H\&K, CH G-band, H$\delta$ and MgB), allowing us to accurately
probe the stellar contents and metallicity gradients, and so
investigate the stellar ages and populations produced in these extreme
systems.  Such observations will allow us to probe further the
timescales involved in the encounters as well as how and when these
galaxies form most of their stars and thus the precise relation to
similarly massive galaxies at the present day: luminous ellipticals.

\section*{acknowledgments}
We thank the anonymous referee for comments and suggestions which
significantly improved the presentation and content of this paper.  We
would also like to thank Jim Geach for providing the {\it HST} image of
SSA\,22.96 and Dave Alexander, Craig Booth, Richard Bower, Alastair
Edge, Vince Eke, Reinhard Genzel, Cedric Lacey, Ian McCarthy, Michael
Merrifield, Chris Mihos and David Wake for useful discussions and help.
We thank Thor Wold, Tim Carroll and Andy Adamson for their excellent
support at UKIRT and James Turner for excellent support and advice at
Gemini-S.  AMS acknowledges support from a PPARC PDRF, IRS acknowledges
support from the Royal Society, AWB acknowledges support the Research
Corporation and the Alfred Sloan Foundation.

\bibliographystyle{apj}
\bibliography{/home/ams/Projects/ref}
%\bibliography{/Users/markswinbank/work/Projects/ref}

%ZZZ new references - FJ, Kovacs et al., Hainline other?

\end{document}